\documentclass[showpacs,twocolumn,aps,prb]{revtex4-1}
\usepackage{amssymb}
\usepackage{amsmath}
\usepackage{graphicx}
\usepackage{subfigure}
\usepackage{color}
\usepackage{makecell}
\usepackage{ulem}

\begin{document}
\title{Electric-field controlled nonlinear anomalous Nernst effect in two-dimensional time-reversal symmetric systems}
    \author{Ying-Li Wu$^{1}$}
	\author{Xiao-Qin Yu$^{1}$}
	\email{yuxiaoqin@hnu.edu.cn}
	\affiliation{$^{1}$ School of Physics and Electronics, Hunan University, Changsha 410082, China}
\begin{abstract}
  It's established that the nonlinear anomalous Nernst effect (NANE), originating from Berry curvature near the Fermi energy,  
  {\color{blue}{is symmetry-permitted only when the crystal symmetry in the transport plane of two-dimensional (2D) materials does not exceed a single mirror line.}}
  Here, we show that an applied direct electric field can lift this symmetry constraint, enabling an electric-field-induced NANE emerge in time-reversal symmetric 2D systems with higher crystallographic symmetries. This electric-field-induced NANE originates from two contributions: the Berry connection polarization, which roots in the electric-field-corrected Berry curvature, and the  anomalous-velocity-modified nonequilibrium Fermi distribution function. Additionally, we propose an alternating temperature gradient as a driving force instead of the conventional steady one, ensuring experimental detection of NANE via second-harmonic measurement techniques.  The behaviour of electric-field-induced NANE in the monolayer graphene has been theoretically and systematically investigated.
\end{abstract}

\maketitle
 \section{Introduction}

 The Hall (Nernst) effects, referring to a generation of transverse current in response to a longitudinal electric field (temperature gradient), represent important paradigms in condensed matter physics\cite{Z. Qiao2010,G. Xu2011,D. Xiao2006}. Recently, the Nernst effect has attracted significant attention for its potential applications in energy harvesting and cooling \cite{Uchida-2021,Uchida-2022}, since the transverse thermoelectric devices based on Nernst effect exhibit several advantages over the longitudinal counterparts, including simple lateral structure, convenient scalability, and easy fabrication. Traditionally, Hall and Nernst effects (e.g., conventional, anomalous, and thermal Hall /Nernst effects) were believed to require  time-reversal ($\mathcal{T}$) symmetry broken. However,
 the 2015 discovery of the nonlinear anomalous Hall effect (NAHE)\cite{I. Sodemann2015}, a second-harmonic response to an alternating electric field, in $\mathcal{T}$-symmetric but inversion ($\mathcal{P}$) symmetric broken systems reveals that Hall voltage can emerge in the $\mathcal{T}$-symmetric systems, reshaping our understanding that Hall responses require the $\mathcal{T}$ symmetry breaking.  NAHE originates from the Berry curvature dipole (BCD)\cite{I. Sodemann2015,T.Low2015,Z.Z.Du2018,Z.Du2019,R.Battilomo2019,J.I.Facio2018,Z.Du2021,A.Bandyopadhyay2024}, a first-order moment of the Berry curvature over the occupied states in momentum space, instead of Berry curvature, attracting broad interests in exploration of the nonlinear transport phenomena stemming from other quantum geometric quantities beyond Berry curvature.  A series of novel nonlinear transport phenomena are identified, such as nonreciprocal magnetoresistance (NMTR)\cite{W.Miao2023,D. Kaplan2024}, intrinsic planar Hall effect(IPHE)\cite{V.A.Zyuzin2020,Hui Wang2024}, nonlinear thermal Hall effect\cite{C. Zeng2020,D.-K. Zhou2022,H.Varshney2,H. Varshney2023,Zhang-2025}, and nonlinear anomalous Nernst effect (NANE)\cite{Zhang-2025,X.-Q. Yu2019,C. Zeng2019,Y.-L. Wu2021,Harsh Varshney2025,Liu-2025}.

NANE\cite{X.-Q. Yu2019,C. Zeng2019,Y.-L. Wu2021}, a nonlinear thermoelectric phenomenon in second-order response to the temperature gradient, originates from Berry curvature near the Fermi energy and has a quantum origin rooted in a pseudotensorial quantity $\Lambda_{cd}^{T}=-\int[d\boldsymbol{k}]\Omega_{c}(\boldsymbol{k})\partial_{k_{d}}
f_{0}(\varepsilon_{\boldsymbol{k}}-E_{f})^{2}/T^{2}$, which is analogous to the BCD, where $f_{0}$ is the equilibrium Fermi distribution function, $\varepsilon_{\boldsymbol{k}}$ represents energy band, and $E_{f}$ indicates Fermi energy.
Like BCD-induced NAHE, this NANE exhibits a linear dependence on relaxation time $\tau$ and require $\mathcal{P}$ symmetry breaking rather than $\mathcal{T}$ symmetry breaking.  In addition, the largest symmetry for nonvanishing NANE in the transport plane of the two-dimensional (2D) materials is a single mirror line\cite{X.-Q. Yu2019,C. Zeng2019,Y.-L. Wu2021}.

Beyond the $\tau$-linearly-dependent NANE originating from quantity ${\mathbf{\Lambda}}^{T}$, an intrinsic $\tau$-independent NANE can also emerge in $\mathcal{PT}$-symmetric systems arising from either the orbital magnetic quadrupole moment \cite{Y. Gao2018} or thermal Berry connection polarizability \cite{Zhang-2025}.  Being analogous to the linear Nernst effect, however, the intrinsic NANE disappears in the presence of $\mathcal{T}$ symmetry.  Consequently, both intrinsic and ${\mathbf{\Lambda}}^{T}$-induced inherent NANE are suppressed in the $\mathcal{T}$-symmetric 2D materials with high crystal symmetries (e.g., those with symmetries exceeding a single mirror line in the transport plane), which severely restricts the selection range of materials for the NANE.

Recently, it has been reported that the symmetry constraint on the BCD-induced NAHE can be lifted via electric-field engineering\cite{X.-G. Ye2023,A. Bhattacharya2025,A.Mukherjee2025,S. Korrapati2025,H.Li2023,J.Yang2025}, wherein an additional direct (dc) electric field $\boldsymbol{E}^\text{dc}$ applied to the NAHE experimental setup induces a field-induced correction to BCD that is proportional to the Berry connection polarization (BCP)\cite{X.-G. Ye2023,A. Bhattacharya2025,A.Mukherjee2025,S. Korrapati2025,H.Li2023,Tanay-2021}.
Importantly, studies reveal that the field-induced BCD can exist in systems where  the inherent BCD and Berry curvature are symmetry-forbidden, including 1) $\mathcal{T}$-symmetric 2D materials with crystal symmetries beyond single mirror symmetry (e.g., dual mirror Rashba systems\cite{A. Bhattacharya2025} and  $C_{2}$-symmetric $p$-wave antiferromagnets\cite{S. Korrapati2025}); and 2) $\mathcal{T}$ symmetry-broken materials, such as $g$-wave altermagnet with four-fold rotation symmetry ($C_{4}$)\cite{S. Korrapati2025} and  $d$-wave altermagnet with combined $C_{4}\mathcal{T}$ symmetry \cite{A.Mukherjee2025,S. Korrapati2025}.
 Consequently, a second-harmonic Hall voltage can be generated from the field-induced BCD in systems where the inherent-BCD-induced NAHE is symmetry-suppressed. Experimentally, the field-induced BCD and its associated NAHE have been observed in thick $T_{d}$-$\mathrm{WTe}_{2}$ with effective $\mathcal{P}$ symmetry in the $x$-$y$ plane\cite{X.-G. Ye2023}.

\begin{figure}[htbp]
\centering
\flushleft
\includegraphics[width=1.0\linewidth,clip]{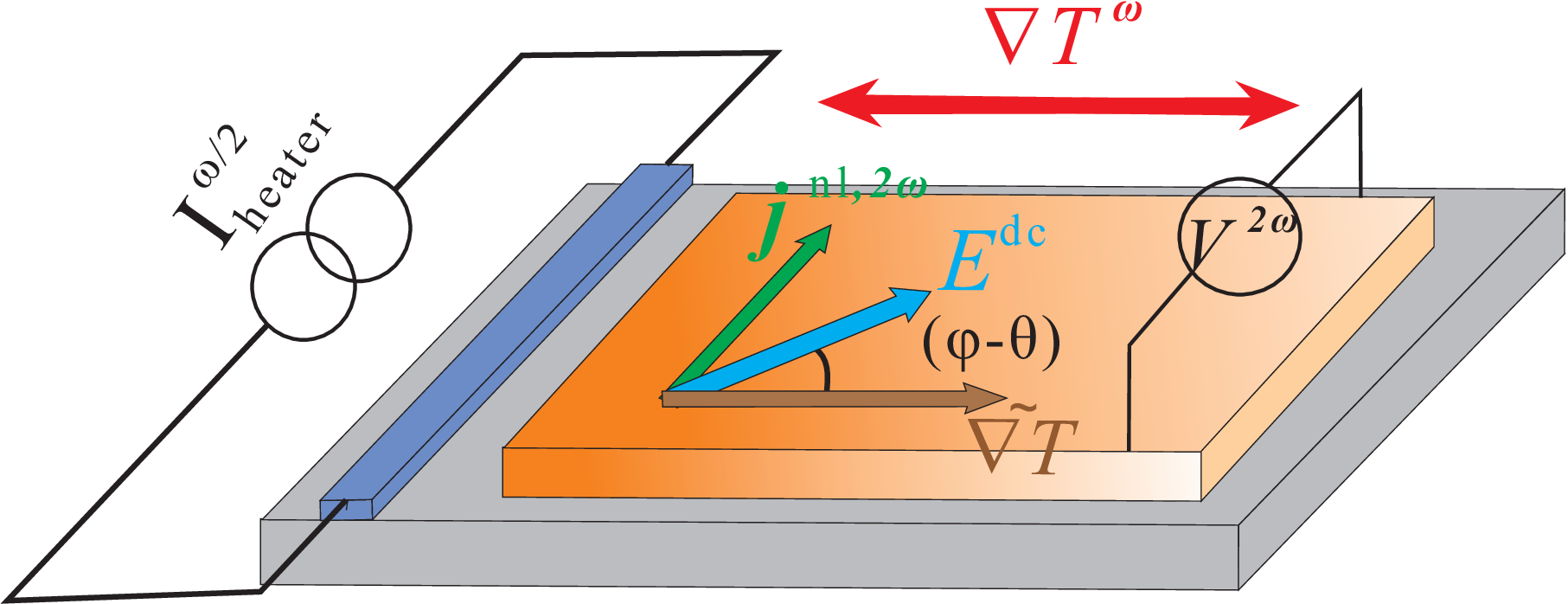}
\caption{Schematic of the measurement setup for electric-field-induced NANE, showing the complex amplitude vector $\tilde{\boldsymbol{\nabla}T}$ of the alternating temperature gradient $\boldsymbol{\nabla} T^{\omega}$ and direct electric field $\boldsymbol{E}^\text{dc}$ in the same plane. 
 $\boldsymbol{\nabla} T^{\omega}$ is generated by passing an alternating current $I^{\omega/2}_\text{heater}$ with frequency $\omega/2$ to a microheater electrode. $\theta$ ($\varphi$)  represents the polar angle of $\tilde{\boldsymbol{\nabla}T}$ ($\boldsymbol{E}^\text{dc}$) with respect to $x$-axis, which is taken as a crystal axis (e.g. the zigzag direction in graphene).
}
\label{figure-sch}
\end{figure}

In this paper, we theoretically investigate the electric-field-induced NANE as a second-harmonic response to an \textit{alternating} temperature gradient $\boldsymbol{\nabla} T^{\omega}(t)=\mathrm{Re} \{\tilde{\boldsymbol{\nabla}T} e^{i\omega t}\}$ (where  $\tilde{\boldsymbol{\nabla}T}\in \mathcal{C}$ 
 is the spatially uniform complex amplitude vector of $\boldsymbol{\nabla} T^{\omega}$, and  $\omega$ indicates frequency) [Fig.~\ref{figure-sch}]  and analyze the crystal symmetry constraint on this NANE in the $\mathcal{T}$-symmetric 2D materials. In previous theoretical studies on NANE or other thermally driven thermoelectric phenomena, researchers typically consider a steady (time-independent) dc temperature gradient $\boldsymbol{\nabla} T^\text{dc}$ and investigate the dc response, as the experimental realization of $\boldsymbol{\nabla} T^{\omega}$ is challenging. However, a regulatable $\boldsymbol{\nabla} T^{\omega}$ has recently been realized experimentally by passing an alternating current $I^{\omega/2}_{h}=I_{0}\sin(\omega t/2+\phi_{0})$ 
 to a microheater electrode\cite{Liu-2025}. The use of alternating temperature gradients in electric-field-induced NANE allows exclusion of spurious nonlinear currents (e.g., second-order in $\boldsymbol{E}^\text{dc}$ and first-order in $\mathrm{Re} \{\tilde{\boldsymbol{\nabla}T}\}$) through measuring the second-harmonic voltage via lock-in amplification (see details in Appendix \ref{EIFO}). We demonstrate the electric-field-induced NANE can emerge in time-reversal symmetric 2D systems with higher crystallographic symmetries. This finding broadens the material platform for generating transverse thermoelectric effects and provides an alternative approach for thermoelectric energy harvesting and cooling applications. In addition to the BCP contribution, we also reveal an extra contribution to the electric-field-induced NANE arising from anomalous-velocity-modified nonequilibrium distribution function, which is absent in electric-field induced NAHE.

 The paper is organized as follows. We firstly review the ${\mathbf{\Lambda}}^{T}$-induced inherent  NANE driven by a steady temperature gradient, and then derive the formula of the rectified ($j_{a}^{\mathrm{nl},0}$) and second-harmonic ($j_{a}^{\mathrm{nl},2\omega}$) currents as a second response to the alternating temperature gradient $\boldsymbol{\nabla} T^{\omega}$ when in the presence of a dc electric field $\boldsymbol{E}^\text{dc}$ based on the Boltzmann theory in Sec.~\ref{TD}. The symmetry constraints on the nonlinear anomalous Nernst coefficients (NANCs) are analyzed in Sec.~\ref{SA-AD}, where the angular dependence of electric-field-induced NANC $\xi_{abcd}$ on the orientations of the electric field and temperature gradient with respect to the principal crystal axis is also investigated. The behavior of the electric-field-induced NANE for the monolayer graphene is discussed in Sec.~\ref{td}. Finally, conclusions are given in Sec.~\ref{fou}.

\section{THEORETICAL REVIEW AND DERIVATION} \label{TD}
The relation between nonlinear anomalous Nernst current $\boldsymbol{j}^{\mathrm{nl}}$ (where the superscript ``nl'' refers to nonlinear) and the pseudotensorial quantity $\Lambda_{cd}^{T}$ has been recently determined through the semiclassical framework of the electron dynamics\cite{X.-Q. Yu2019,C. Zeng2019}. We here briefly review the NANE in 2D systems, in which the Berry curvature transforms from a pseudovector to a pseudoscalar, with only the out-of-plane component ($\Omega_c=z$) remaining nonzero. Consequently, $\Lambda_{cd}^{T}$ exhibits pseudovector behavior confined to the 2D plane as
 \begin{equation}
\begin{aligned}
\mathbf{\Lambda}^{T}=-\int[d\boldsymbol{k}]\frac{(
\varepsilon_{\boldsymbol{k}}-E_{f})^{2}}{T^{2}}
\frac{\partial f_{0}}{\partial \boldsymbol{k}}\Omega_{z}(\boldsymbol{k}),
\end{aligned}
\end{equation}
where $\int[d\boldsymbol{k}]$ is the shorthand for $\int d\boldsymbol{k}/(2\pi)^{2}$,
 $\Omega_{c}(\boldsymbol{k})=[\nabla_{\boldsymbol{k}}\times\boldsymbol{\mathcal{A}}
 (\boldsymbol{k})]_{c}$ denotes the inherent Berry curvature along the $c$ direction,
and $\boldsymbol{\mathcal{A}}(\boldsymbol{k})=\langle u_{n}(\boldsymbol{k})|i\nabla_{\boldsymbol{k}}|u_{n}(\boldsymbol{k})\rangle$ denotes the intraband Berry
connection with $|u_{n}(\boldsymbol{k})\rangle$ being the periodic part of the Bloch wave-function for band $n$. In fact, the largest symmetry allowed for a nonvanishing $\mathbf{\Lambda}^{T}$ in the transport plane ($x$-$y$ plane) is a single mirror line (a mirror plane that is orthogonal to the 2D crystal)\cite{X.-Q. Yu2019}. 
The nonlinear Nernst current $\boldsymbol{j}^{\mathrm{nl}}$ as a response to the second order in steady dc temperature gradient $\boldsymbol{\nabla} T^\text{dc}$  is
\begin{equation}
\begin{aligned}
\boldsymbol{j}^{\mathrm{nl}}=\frac{e\tau}{\hbar^{2}}(\hat{z}\times\boldsymbol{\nabla} T^\text{dc})(\boldsymbol{\nabla} T^\text{dc}\cdot\mathbf{\Lambda}^{T}),
\end{aligned}
\label{fd}
\end{equation}
where $\hbar$ is the Planck constant, and $\tau$ refers to relaxation time.

Within the extended semiclassical theory, it has been established that Berry connection experiences a gauge-invariant correction $\mathcal{A}_{a}^{E}=\tilde{\mathcal{G}}_{ab}(\boldsymbol{k})E^{\text{dc}}_{b}$ under an applied $\boldsymbol{E}^\text{dc}$,\cite{Y. Gao2014,S. Lai2021,H.Liu2021,H. Liu2022} where $\tilde{\mathcal{G}}_{ab}(\boldsymbol{k})$ represents the component of BCP tensor and is defined for the band $n$ as:
\begin{equation}
\begin{aligned}
&\tilde{\mathcal{G}}_{ab}(\boldsymbol{k})=2e\mathrm{Re}\sum_{m\neq n}
\frac{\mathcal{A}_{a}^{nm}(\boldsymbol{k})\mathcal{A}_{b}^{mn}(\boldsymbol{k})}
{  \varepsilon^{n}_{\boldsymbol{k}}-\varepsilon^{m}_{\boldsymbol{k}}},
\end{aligned}
\label{Gmaj}
\end{equation}
with $\mathcal{A}_{a}^{nm}(\boldsymbol{k})=\langle u_{n}(\boldsymbol{k})|i \partial_{k_{a}}|u_{m}(\boldsymbol{k})\rangle$ indicating the interband Berry connection. This gauge-invariant correction directly
 generates a field-induced Berry curvature $\boldsymbol{\Omega}^{E}(\boldsymbol{k})=\nabla\times\boldsymbol{\mathcal{A}}^{E}(\boldsymbol{k})$,
 a correction to the Berry curvature as $\boldsymbol{\tilde{\Omega}}\left(\boldsymbol{k}\right)=\boldsymbol{\Omega}(\boldsymbol{k})+
 \boldsymbol{\Omega}^{E}(\boldsymbol{k})$, which can lead to a electric-field-induced NANE investigated as following.

 \begin{table*}[tbph]
\centering
\caption{Symmetry constraints on $\xi_{abcd}^{\mathrm{BCP}/\mathrm{NDF-AV}}$ and $\chi_{abc}$ tensors from point group symmetries pertaining to 2D time-reversal symmetric  materials. $'\checkmark'$ ($'\times'$) means the element is symmetry allowed (forbidden).
Here, $M_{b}$, $C^{b}_{n}$, and $S^{b}_{n}$ represent mirror, $n$-fold rotation, and $n$-fold rotation-reflection symmetry operation along the $b$ direction for $b=\c\{x,y,z\}$, respectively. 
}
\begin{centering}
\begin{tabular*}{18 cm}{@{\extracolsep{\fill}}c cccc ccc |cccc}
\hline \hline
\multicolumn{1}{c} { }   & \multicolumn{7}{c|} {Elemental crystallographic symmetries}     &\multicolumn{4}{c} {Point Groups  }\\
\cline{2-12}
  \thead{}
  &\thead{$C_{2}^{z}$ }        &\thead{$C_{2}^{x}$ }    &\thead{$C_{2}^{y}$  }
  &\thead{$C_{3,4,6}^{z}$\\$S_{4,6}^{z}$}   & \thead{$M_{z}$}         &\thead{$M_{x}$  }            &\thead{$ M_{y}$}
  &\thead{$C_{1v}(\sigma_{y})$}
  &\thead{$C_{1v}(\sigma_{x})$}
  &\thead{$C_{2v}$, $D_{2}$}             &\thead{ $C_{3v,4v,6v}$, $D_{3,4,6}$\\$D_{2d}$, $D_{3h}$} \\
 \hline
 $\xi_{xyyy}^{\mathrm{BCP}}$       &$\checkmark$            &$\times$          &$\times$
  &$\checkmark$                    &$\checkmark$            & $\times$         &$\times$
  & $\times$                       &$\times$                &$\times$          &$\times$  \\
 $\xi_{xyyx}^{\mathrm{BCP}/\mathrm{NDF-AV}}$   &$\checkmark$     &$\checkmark$      &$\checkmark$
 &$\checkmark$                   &$\checkmark$              &$\checkmark$      &$\checkmark$
 &$\checkmark$                   &$\checkmark$              &$\checkmark$      &$\checkmark$ \\
 $\xi_{yxxx}^{\mathrm{BCP}}$     &$\checkmark$              &$\times$          &$\times$
 &-$\xi_{xyyy}^{\mathrm{BCP}} $  &$\checkmark$              & $\times$         &$\times$
 & $\times$                      &$\times$                  & $\times$         & $\times$\\
 $\xi_{yxxy}^{\mathrm{BCP}/\mathrm{NDF-AV}}$                    &$\checkmark$      &$\checkmark$                  &$\checkmark$                   &$\xi_{xyyx}^{\mathrm{BCP}/\mathrm{NDF-AV}}$      & $\checkmark$                   &$\checkmark$                   &$\checkmark$              &$\checkmark$      & $\checkmark$         &$\checkmark$                   &$\xi_{xyyx}^{\mathrm{BCP}/\mathrm{NDF-AV}}$                              \\
  $\chi_{xyy}$                   &$\times$                 &$\checkmark$      &$\times$
 &$\times$                       &$\checkmark$             &$\times$          &$\checkmark$
 &$\checkmark$                   &$\times$                 &$\times$           &$\times$ \\
 $\chi_{yxx}$                    &$\times$                 &$\times$          &$\checkmark$
  &$\times$                      &$\checkmark$             &$\checkmark$      &$\times$
  &$\times$                      &$\checkmark$             &$\times$           & $\times$ \\
 \hline \hline
\end{tabular*}
\par\end{centering}
\label{table1}
\end{table*}

Taking the Berry-phase correction to the orbital magnetization into consideration, the formula of  charge transport current $\boldsymbol{j}$ driven by the statistical force, arising from the the steady temperature gradient $\boldsymbol{\nabla} T$, has been derived through the semiclassical theory in previous works \cite{D. Xiao2006,X.-Q. Yu2019} and is given by
\begin{equation}
\begin{aligned}
\boldsymbol{j}
=&-e\int[d\boldsymbol{k}]\dot{\boldsymbol{r}}f-{\frac{\boldsymbol{\nabla} T}{T}}\times\frac{e}{\hbar}
\int[d\boldsymbol{k}]\mathbf{\tilde{\Omega}}(\boldsymbol{k})
\left[(\varepsilon_{\boldsymbol{k}}-\mu)f\left(\boldsymbol{k}\right)\right. \\
&\left.+k_{B}T\mathrm{ln}(1+e^{-\frac{(\varepsilon_{\boldsymbol{k}}-\mu)}{k_\text{B}T}}) \right]
\label{curre0},
\end{aligned}
\end{equation}
where the first term is conventional current, and the second term is anomalous charge current stemming from the Berry curvature, $f\left(\boldsymbol{k}\right)$ represents the nonequilibrium distribution function (NDF), which was initially written as equilibrium distribution function $f_{0}$ in the original work\cite{D. Xiao2006} but has been extended into the nonequilibrium situation by Yu \textit{et al.\cite{X.-Q. Yu2019}} Considering the effect of the electric-field-induced Berry curvature $\boldsymbol{\Omega}^{E}(\boldsymbol{k})$, we have replaced the inherent Berry curvature $\boldsymbol{\Omega}(\boldsymbol{k})$ by the electric-field corrected Berry curvature $\boldsymbol{\tilde{\Omega}}(\boldsymbol{k})$ in Eq.~(\ref{curre0}).

When in presence of an alternating temperature gradient $\boldsymbol{\nabla} {T}^{\omega}$ and a dc electric field $\boldsymbol{E}^\text{dc}$ simultaneously, $f\left(\boldsymbol{k}\right)$ can be determined through solving the Boltzmann transport equation. We are interested in the response up to the second order in temperature gradient and first order in electric field, and hence have approximated NDF as  $f\approx\mathrm{Re}\{f_{0}+\delta f^{(1,0)}+\delta f^{(0,1)}+\delta f^{(1,1)}+\delta f^{(0,2)}\}$ with the high-order terms  $\delta f^{(n,m)}$, which are $n$-th order in $E^{\text{dc}}$ and $m$-th order in $\text{Re}\{\tilde{\partial_{a}{T}}\}$, understood to vanish. After a series of careful derivation (see details in Appendix \ref{DF-T}), the formulas for expansion coefficients [$\delta f^{(1,0)}$, $\delta f^{(0,1)}$, $\delta f^{(1,1)}$ and $\delta f^{(0,2)}$] can be determined as
\begin{equation}
\begin{aligned}
&\delta f^{(1,0)}=\frac{e\tau}{\hbar}\boldsymbol{E}^{\text{dc}}\cdot\nabla_{\boldsymbol{k}}f_{0},~~~~~
\delta f^{(0,1)}=\delta f^{(0,1)}_{\omega}e^{i\omega t},\\
&\delta f^{(1,1)}=\delta f^{(1,1)}_{\omega}e^{i\omega t},~~~~\delta f^{(0,2)}=\delta f^{(0,2)}_{0}+\delta f^{(0,2)}_{2\omega}e^{2i\omega t},\\
\end{aligned}
\label{a-f}
\end{equation}
where the amplitudes [$\delta f^{(0,1)}_{\omega}$, $\delta f^{(1,1)}_{\omega}$, $\delta f^{(0,2)}_{0}$, and $\delta f^{(0,2)}_{2\omega}$ ] are given in Eq.~(\ref{Ap-A-11}).
Hence, the $a$-component of nonlinear anomalous Nernst current  $j_{a}^{\mathrm{nl}} $ as second-order response to the temperature gradient can be written as (see the details in Appendix \ref{ESNAE})
 \begin{equation}
 {j}_{a}^{\mathrm{nl}}=\mathrm{Re}\{{j}_{a}^{ {\mathrm{nl}},0}+j_{a}^{\mathrm{nl},2\omega}e^{2i\omega t}\},
 \end{equation}
where
${j}_{a}^{ {\mathrm{nl}},0}$ and $j_{a}^{\mathrm{nl},2\omega}$ represents the rectified and the second harmonic current, respectively, and can be written as
\begin{equation}
\begin{aligned}
j_{a}^{\mathrm{nl},0}&=\left(\chi_{abc}+\xi_{abcd}E^{\text{dc}}_{d}\right)(\tilde{\partial_{b}T})^{*}\tilde{\partial_{c}{T}},\\
j_{a}^{\mathrm{nl},2\omega}&=\left(\chi_{abc}+\xi_{abcd}E^{\text{dc}}_{d}\right)\tilde{\partial_{b}
{T}}\tilde{\partial_{c}{T}}.
\end{aligned}
\label{jmjnl}
\end{equation}
After a series of careful derivation (details can be found in Appendix \ref{ESNAE}), the explicit expressions for the inherent ($\chi_{abc}$) and electric-field-induced ($\xi_{abcd}$) NANCs have been systematically derived for both 2D and three-dimensional systems with $\mathcal{T}$ symmetry .  We focus exclusively on the 2D systems. 
The derived forms of $\chi_{abc}$ and $\xi_{abcd}$ in $\mathcal{T}$-symmetric 2D systems are given by
\begin{equation}
\begin{aligned}
\chi_{abc}&=\frac{\tau e\epsilon_{abz}}{2(1+i\omega\tau)\hbar^{2}}\Lambda^{T}_{c},~~~~\xi_{abcd}=\xi^{\text{BCP}}_{abcd}
+\xi^{\text{NDF-AV}}_{abcd},\\
\xi^\text{BCP}_{abcd}&=\frac{\tau e\epsilon_{abz}}{2(1+i\omega\tau)\hbar^{2}}\Upsilon_{cd},~~~
\xi^\text{NDF-AV}_{abcd}=\frac{\tau e^{2}\epsilon_{abz}\epsilon_{cdz}}{2(1+i\omega\tau)\hbar^{2}}
\Gamma,
\label{c-chi}
\end{aligned}
\end{equation}
where the superscript ``AV'' refer to anomalous velocity, $\epsilon_{abz}$ represents the Levi-Civita symbol, and pseudotensorial (scalar) quantity $\Upsilon_{cd}$ $(\Gamma)$ is defined as
\begin{equation}
\begin{aligned}
\Upsilon_{cd}&=\int[d\boldsymbol{k}]
\left[\frac{\partial{\tilde{\mathcal{G}}_{xd}\left(\boldsymbol{k}\right)}}{\partial k_{y}}
-\frac{\partial{\tilde{\mathcal{G}}_{yd}\left(\boldsymbol{k}\right)}}{\partial k_{x}}\right]
     \frac{(\varepsilon_{\boldsymbol{k}}-E_{f})^{2}}{T^{2}}\frac{\partial f_{0}}{\partial k_{c}},\\
\Gamma&=-
\int[d\boldsymbol{k}] \Omega^{2}_{z}(\boldsymbol{k}) \frac{(\varepsilon_{\boldsymbol{k}}-E_{f})^{2}}{T^{2}}\frac{\partial f_{0}}{\partial \varepsilon_{\boldsymbol{k}}}.
\end{aligned}
\label{coeffu}
\end{equation}

The component $\xi^\text{BCP}_{abcd}$ stems from BCP, while  $\xi^\text{NDF-AV}_{abcd}$ arises from the NDF modified by the anomalous velocity ($\boldsymbol{E}\times\mathbf{\Omega}$). The BCP contribution roots in the electric-field-induced Berry curvature. In this work, we focus on the second-harmonic response, which can be measured via lock-in amplification. Hence, only  $j_{a}^{\mathrm{nl},2\omega}$ will be investigated in subsequent analysis. In vector notation, the second harmonic current $\boldsymbol{j}^{\mathrm{nl},2\omega}$ can be further written as
\begin{equation}
\begin{aligned}
\boldsymbol{j}^{\mathrm{nl},2\omega}&=
\frac{e\tau}{2(1+i\omega\tau)\hbar^{2}}\left(\hat{z}\times\tilde{\boldsymbol{\nabla}T}\right)
\left(\tilde{\boldsymbol{\nabla}{T}}\cdot\mathbf{\Lambda}^{T}\right)\\
& +\frac{e\tau}{2(1+i\omega\tau)\hbar^{2}}\left(\hat{z}\times\tilde{\boldsymbol{\nabla} {T}}\right)\left(\tilde{\boldsymbol{\nabla} {T}}\cdot \overleftrightarrow{\boldsymbol{\Upsilon}}\right)\cdot \boldsymbol{E}^\text{dc}\\
&+\frac{e^{2}\tau \Gamma}{(1+i\omega\tau)\hbar^{2}}\left(\hat{z}\times\tilde{\boldsymbol{\nabla} {T}}\right)\left(\hat{z}\times\tilde{\boldsymbol{\nabla} {T}}\right)\cdot \boldsymbol{E}^\text{dc},\\
\end{aligned}
\label{totalcurren}
\end{equation}
showing the current $\boldsymbol{j}^{\mathrm{nl},2\omega}$ corresponds to Nernst current, as $\boldsymbol{j}^{\mathrm{nl},2\omega}\cdot \tilde{\boldsymbol{\nabla}T}=0$. When the dc electric field $\boldsymbol{E}^{\text{dc}}$ is aligned with  the complex amplitude vector of temperature gradient $\tilde{\boldsymbol{\nabla}T}$ (i.e. $\boldsymbol{E}^{\text{dc}}\parallel\tilde{\boldsymbol{\nabla}T}$), the third terms originating from $\xi^\text{NDF-AV}_{abcd}$ in Eq.~{\eqref{totalcurren}} becomes zero, indicating that the contribution from the anomalous-velocity-modified NDF vanishes, and the electric-field-induced NANE solely stems from the BCP.

\section{SYMMETRIES ANALYSIS and angular dependency} \label{SA-AD}
Equation~\eqref{c-chi} implies both the rank-three tensor $\chi_{abc}$ and the rank-four tensor {\color{blue}{$\xi_{abcd}$}} are antisymmetric in $a$ and $b$, namely $\chi_{abc}=-\chi_{bac}$ and $\xi_{abcd}=-\xi_{bacd}$. This antisymmetry implies all components with $a=b$ are zero since $\chi_{aac}=-\chi_{aac}$ and  $\xi_{aacd}=-\xi_{aacd}$. As a result, there are at most two (four) independent tensor elements for $\chi$ ($\xi$) in the 2D systems, which we choose here as $[\chi_{xyy},~\chi_{yxx}$] and [$\xi_{xyyx},~\xi_{xyyy},~\xi_{yxxx},~\xi_{yxxy}$]. Furthermore, we note $\xi^\text{NDF-AV}_{xyyy}=0$ and $\xi^\text{NDF-AV}_{yxxx}=0$ since the contribution from the anomalous-velocity-modified NDF to $\boldsymbol{j}^{\mathrm{nl},2\omega}$ [i.e., the third term in Eq. \eqref{totalcurren}] vanishes when $\boldsymbol{E}^{\text{dc}}\parallel\tilde{\boldsymbol{\nabla}T}$.


Further constraints on the coefficients [$\chi_{abc}$, and $\xi_{abcd}$] come from the crystalline point-group symmetries including $\mathcal{P}$ symmetry. Under $\mathcal{P}$ symmetry, the current $\boldsymbol{j}^{\mathrm{nl}}$ changes sign but $\tilde{\partial_{b}T}\tilde{\partial_{c}T}$ remains unchanged, restricting $\chi_{abc}$ $\mathcal{P}$-odd ($\chi_{abc}\xrightarrow{\mathcal{P}}-\chi_{abc}$). This $\mathcal{P}$-odd parity forces $\chi_{abc}$ to vanish in the systems with $\mathcal{P}$ symmetry. However, the $d$-component of dc electric field $E^{\mathrm{dc}}_{d}$ [Eq.~\eqref{jmjnl}] also changes sign under $\mathcal{P}$ symmetry, resulting in $\xi_{abcd}$ $\mathcal{P}$-even ($\xi_{abcd}\xrightarrow{\mathcal{P}}\xi_{abcd}$). The coefficient $\xi_{abcd}$ has been decomposed into two parts as $\xi_{abcd}=\xi^\text{BCP}_{abcd}
+\xi^\text{NDF-AV}_{abcd}$ [Eq.~\eqref{c-chi}], 
and one can further confirm $\xi^\text{NDF-AV}_{abcd}=0$ in presence of both $\mathcal{T}$ and $\mathcal{P}$ symmetries. That's because Berry curvature satisfies $\boldsymbol{\Omega}(\boldsymbol{k})=-\boldsymbol{\Omega}(-\boldsymbol{k})$ under $\mathcal{T}$ symmetry and $\boldsymbol{\Omega}(\boldsymbol{k})=\boldsymbol{\Omega}(-\boldsymbol{k})$ under $\mathcal{P}$ symmetry, resulting in $\boldsymbol{\Omega}(\boldsymbol{k})=0$ at each point in the momentum space when both $\mathcal{T}$ and $\mathcal{P}$ symmetries are present. Be different to the Berry curvature, BCP $\mathcal{\tilde{G}}\left(\boldsymbol{k}\right)$ can exist in the presence of both $\mathcal{P}$ and $\mathcal{T}$ symmetries since the sign of $\mathcal{\tilde{G}}\left(\boldsymbol{k}\right)$ remains unchange under $\mathcal{T}$/$\mathcal{P}$ symmetries, namely $\mathcal{\tilde{G}}\left(\boldsymbol{k}\right)\xrightarrow{\mathcal{P}/\mathcal{T}}\mathcal{\tilde{G}}
\left(-\boldsymbol{k}\right)$. Consequently, a nonlinear Nernst current can also emerge in the systems with both $\mathcal{T}$ and $\mathcal{P}$ symmetries, which attributes to the nonzero electric-field-induced NANC from BCP.

In addition to $\mathcal{T}$ and $\mathcal{P}$ symmetries, the other crystal symmetry operations $\mathcal{R}$ also impose constraint on $\chi_{abc}$ and $\xi_{abcd}$. Since $\chi^{\prime}=\chi/\tau$  ($\xi^{\prime}=\xi/\tau$) is a third-rank (fourth-rank) $\mathcal{T}$-even tensor, the constraint imposed by $\mathcal{R}$ can be expressed as\cite{Robert-2005}
\begin{equation}
\begin{aligned}
&\chi_{abc}=\mathcal{R}_{aa^{\prime}}\mathcal{R}_{bb^{\prime}}\mathcal{R}_{cc^{\prime}}
\chi_{a^{\prime}b^{\prime}c^{\prime}},\\
&\xi_{abcd}=\mathcal{R}_{aa^{\prime}}\mathcal{R}_{bb^{\prime}}
\mathcal{R}_{cc^{\prime}}\mathcal{R}_{dd^{\prime}}\xi_{a^{\prime}b^{\prime}c^{\prime}d^{\prime}},
\end{aligned}
\label{symmety-con}
\end{equation}
where ${\mathcal{R}}$ represents a point group operation, and ${\mathcal{R}_{aa^{\prime}}}$ are the matrix elements of the symmetry operation in
Cartesian coordinate representation. The repeated indices in Eq.~(\ref{symmety-con}) are summed over.
The point group symmetry constraints on the independent coefficients $[\chi_{xyy},~\chi_{yxx},~\xi^\text{NDF-AV}_{xyyx}$, $\xi^\text{NDF-AV}_{yxxy},~\xi^\text{BCP}_{xyyx},
~\xi^\text{BCP}_{xyyy},~\xi^\text{BCP}_{yxxx},~\xi^\text{BCP}_{yxxy}$] for 2D case are determined by Eq. (\ref{symmety-con}) and summarized in Table \ref{table1}.

 In addition to the symmetry operations confined to the transport plane ($x$-$y$ plane), including $n$-th fold rotational symmetries $C^{z}_{n}$ ($n=2,3,4,6$) around $z$ axis, and the vertical mirror symmetries $M_{a}$ $(a={x~\text{or}\,y})$, we have also analyzed three kinds of out-of-plane symmetry operations that map $z$ to $-z$, namely twofold rotation symmetries $C^{a}_{2}$ around $a$ axis, a horizontal mirror symmetry $M_{z}$, and rotation-reflection symmetries $S^{z}_{4}$ and $S^{z}_{6}$. These out-of-plane symmetry operations are permissible in certain layered 2D crystal materials. The ${\mathbf{\Lambda}}^{T}$-induced inherent NANE are permitted in out-of-plane symmetries $C^{a}_{2}$ and $M_{z}$. Obviously, the applied dc electric field $\boldsymbol{E}^\text{dc}$ removes the symmetry constraint on the NANE, specifically eliminating the requirement that the largest symmetry  allowing $\chi_{aa_{\perp}a_{\perp}}\neq0$ in the transport plane is the single mirror symmetry $M_{a_{\perp}}$ ($a_{\perp}$ indicates axis perpendicular to $a=\{x~\text{or}~y\}$ axis in 2D plane)\cite{X.-Q. Yu2019}. 
All the independent NANCs $[\xi^\text{NDF-AV}_{xyyx},~\xi^\text{NDF-AV}_{yxxy}
~\xi^\text{BCP}_{xyyx},~\xi^\text{BCP}_{xyyy},~\xi^\text{BCP}_{yxxx},~\xi^\text{BCP}_{yxxy}]$ controlled by electric field  are permitted in $C^{z}_{2}$-symmetric systems (the effective $\mathcal{P}$ symmetry in the transport plane). When $\boldsymbol{E}^{\text{dc}}\perp \tilde{\boldsymbol{\nabla}T}$, the coefficients $\xi^\text{NDF-AV/BCP}_{aa_{\perp}a_{\perp}a}$
are allowed across all elemental crystallographic symmetries applicable to 2D systems [Table. \ref{table1}]. Conversely, when $\boldsymbol{E}^{\text{dc}}\| \tilde{\boldsymbol{\nabla}T}$, the term $\xi^\text{BCP}_{aa_{\perp}a_{\perp}a_{\perp}}$ is permitted only for the symmetries [$C^{z}_{n=2,3,4,6}$, $S^{z}_{4,6}$ and $M_{z}$] but forbidden by $C^{x,y}_{2}$ and $M_{x,y}$ symmetries.

Having determined the symmetry-allowed components of the response tensors, we now investigated the angular dependence of the nonlinear anomalous Nernst current.
For the most Nernst transport experiments, the setup has a planar geometry, where the temperature gradient and the generated current are both within the plane (labeled as the $x-y$ plane). In this work, we further assume the $\boldsymbol{E}^\text{dc}$, which was applied to control the NANE, coplanar with the alternating temperature gradient ${\boldsymbol{\nabla} {T}^{\omega}}$. In general, ${\boldsymbol{\nabla} {T}^{\omega}}$ and $\boldsymbol{E}^\text{dc}$ are not aligned with the crystal axis but orients in a polar angle with respect to the $x$-direction (having taken to be a  crystal direction), i.e., $(\tilde{\partial_{x}{T}}, \tilde{\partial_{y}{T}})=\tilde
{\partial{T}}(\cos\theta,\sin\theta$)  and $(E_{x},E_{y})=E(\cos\varphi,\sin\varphi)$, where $\tilde
{\partial{T}}=[(\tilde{\partial_{x}{T}})^{2}+(\tilde{\partial_{y}{T}})^{2}]^{1/2}$ denotes the amplitude of the temperature gradient for simplification.
Then, the electric-field-induced nonlinear second-harmonic  Nernst current $j^{\mathrm{nl},2\omega}_\text{N}$ (where the subscript ``{N}" represents the Nernst), flowing  perpendicularly to the temperature gradient $\boldsymbol{\nabla} {T}^{\omega}$, can be expressed as
\begin{equation}
\begin{aligned}
j^{\mathrm{nl},2\omega}_\text{N}&=\xi_{\text{N}}(\theta,\varphi)
\tilde{\partial{T}}\tilde
{\partial{T}}E,
\end{aligned}
\label{CUR-angle}
\end{equation}
with the angle-dependent scalar coefficient $\xi_{\text{N}}(\theta,\varphi)$  determined by the combination of the $\xi$ tensor elements as (see Appendix \ref{D-app} for a detail discussion)
\begin{equation}
\begin{aligned}
\xi_\text{N}(\theta,\varphi)=&(\xi^{\text{BCP}}_{yxxx}\cos\varphi
   +\xi_{yxxy}\sin\varphi)\cos\theta\\
&-(\xi_{xyyx}\cos\varphi
   +\xi^\text{BCP}_{xyyy}\sin\varphi)\sin\theta,\\
\end{aligned}
\label{d-dffd}
\end{equation}
where $\xi_{abcd}=\xi^\text{NDF-AV}_{abcd}+\xi^\text{BCP}_{abcd}$ with $\xi^\text{NDF-AV}_{yxxx}=\xi^\text{NDF-AV}_{xyyy}=0$ (see details in Sec.~\ref{SA-AD}). Under the symmetry constraints outlined in Table \ref{table1} for the tensor components $\xi_{abcd}$, the angular dependence of the electric-field-induced NANE response  becomes  richer and  more interesting. 
For example, the angular dependence of the electric-field-induced NANC reduces to
\begin{equation}
\xi_\text{N}(\theta,\varphi)=\left(\xi^\text{BCP}_{yxxy}+\xi^\text{NDF-AV}_{yxxy}\right)\sin(\varphi-\theta)
 \label{angel}
\end{equation}
for the  groups $C_{nv}$ and dihedral groups  $D_{n}$ ($n>2$), since they support only one independent tensor element $\xi_{xyyx} =\xi_{yxxy}$ [Table.~\ref{table1}]. 

Equation~\eqref{d-dffd} shows that the electric-field-induced NANE can originate from both BCP contribution and anomalous-velocity-modified NDF contribution. However, the contribution from the anomalous-velocity-modified NDF [$\xi^\text{NDF-AV}$] can be actually eliminated by aligning the electric field and temperature gradient  along $\pm x~ \text{or}~\pm y $ axis. This exclusion is achieved through two configurations: 1) Parallel alignment ($\theta,~\varphi=0~\text{or}~\pi$): aligning $\boldsymbol{E}^{\text{dc}}$ and $\boldsymbol{\nabla}T^{\omega}$ along $\pm x$ results in $|\xi_\text{N}|=\xi^\text{BCP}_{yxxx} $, showing that the electric-field-induced NANE originates solely from BCP; 2)Perpendicular alignment ($\theta,~\varphi={\pi}/{2},~{3\pi}/{2}$): aligning fields $\boldsymbol{E}^{\text{dc}}$ and $\boldsymbol{\nabla}T^{\omega}$ to $\pm y$ yields $|\xi_\text{N}|=\xi^\text{BCP}_{xyyy}$, indicating the BCP contribution dominates, whereas the contribution from the anomalous-velocity-modified NDF vanishes. Therefore, the geometric alignment of $\boldsymbol{E}^{\text{dc}}$ and $\boldsymbol{\nabla}T^{\omega}$ eliminates the anomalous-velocity-modified-NDF contribution, confirming BCP as the sole origin of NANE in both cases. 

\section{The electric-field-induced nonlinear anomalous Nernst effect in monolayer graphene }\label{td}
\begin{figure}[htbp]
\centering
\flushleft
\includegraphics[width=1.0\linewidth,clip]{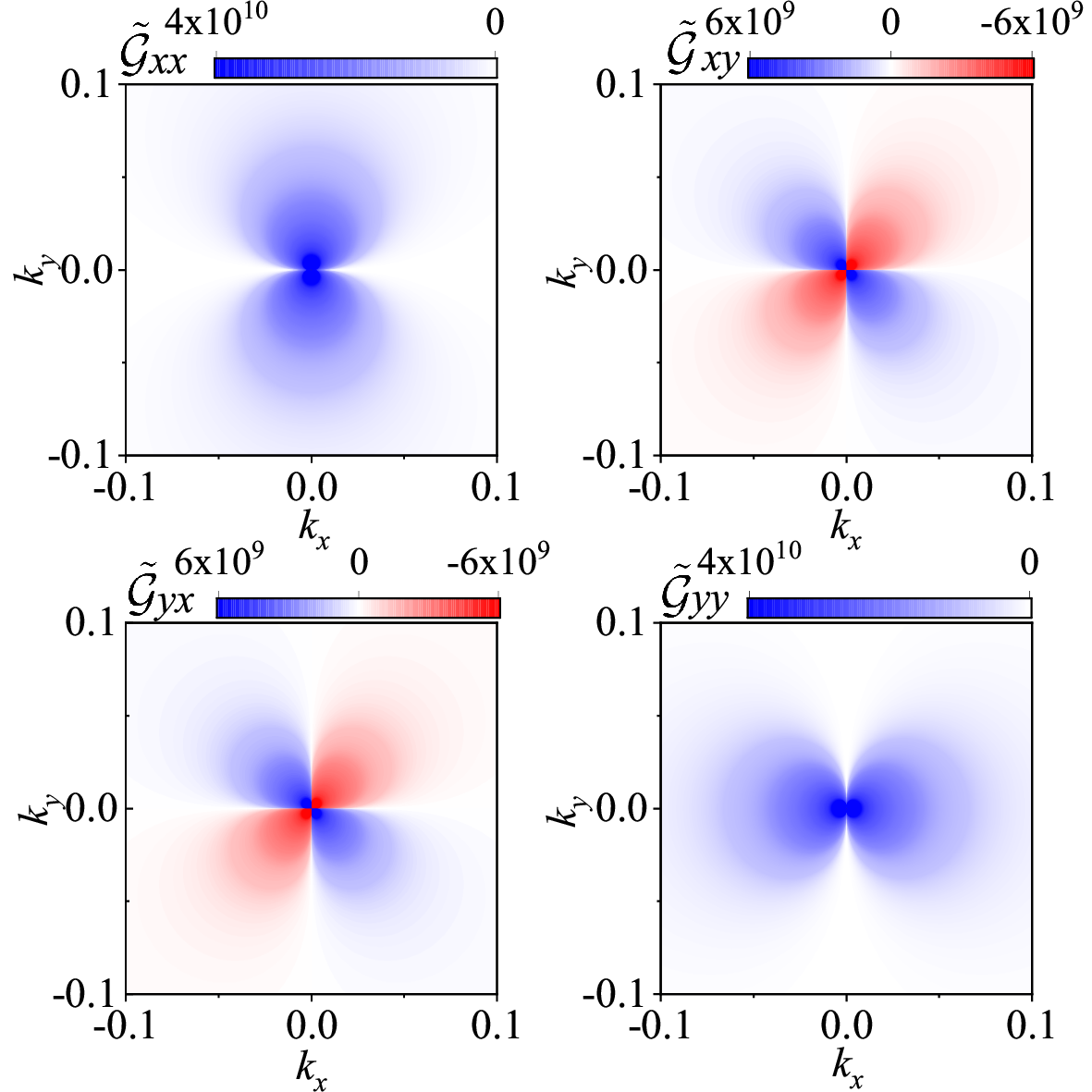}
\caption{Schematic of the Berry connection polarization elements
 $\tilde{\mathcal{G}}_{xx}$ [(a)],  $\tilde{\mathcal{G}}_{xy}$ [(b)], $\tilde{\mathcal{G}}_{yx}$ [(c)] and $\tilde{\mathcal{G}}_{yy}$ [(d)]
 of the conduction band for the $K$ valley of monolayer graphene. 
 Momenta are measured in units of the inverse of the lattice constant $\alpha$ and $\tilde{\mathcal{G}}_{ab}$ is measured in units of $e\alpha^{2}/t_{0}$
  }
\label{fig-BCP}
\end{figure}
The point group symmetry $C_{6v}$ of monolayer graphene includes a threefold rotation $C_{3}$, a twofold rotation $C_{2}$ (an effective $\mathcal{P}$ symmetry in 2D), and the mirror symmetry $M_{x}$. The presence of $C_{3}$ and $C_{2}$ symmetries suppresses the $\mathbf{\Lambda}^{T}$-induced inherent NANE.  Additionally, $\mathcal{T}$ symmetry forces  both linear anomalous Nernst effect and intrinsic NANE to vanish in monolayer graphene. Previously\cite{Y.-L. Wu2021}, we proposed strain and substrate engineering  to break these symmetries and induce NANE in the strained monolayer graphene. 
As shown in Table~\ref{table1}, however, the electric-field-induced NANE are actually permitted under $C_{6v}$ symmetry. Consequently, the NANE can emerge in the pristine monolayer graphene via electric field control, eliminating the need for additional materials engineering.

The low-energy Hamiltonian of the monolayer graphene is \cite{D.Xiao2007,Castro-Neto2009}
\begin{equation}
\begin{aligned}
\hat{H}_{\tau_{v}}(\boldsymbol{k})=
&\tau_{v}v_{F}\hbar k_{x}\sigma_{x}+v_{F}\hbar k_{y}\sigma_{y},
\end{aligned}
\label{ml-Hamiltonian}
\end{equation}
where $x$ axis is selected as the zigzag direction, 
$\tau_{v}=\pm1$ is the valley index, the Fermi velocity $\hbar v_{F}=\sqrt{3}\alpha t_{0}/2$ is determined by the lattice constant $\alpha=0.246~nm$ and the nearest-neighbor hopping $t_{0}\approx2.8~\mathrm{eV}$, and $\sigma$ represents the Pauli matrices for the two basis functions of energy band.
This effective Hamiltonian $\hat{H}_{\tau_{v}}(\boldsymbol{k})$ acts in the space of two-component wave functions $\Phi=(\phi_{A},\phi_{B})$ with $\phi_{\beta(=A,B)}$ representing the electron amplitude on sublattices $\beta$.  In this space, we have the group representation as 
 $\hat{\mathcal{T}}=\sigma_{0}\mathcal{K}$, $\hat{\mathcal{C}_{2}}=\sigma_{x}$, $\hat{M}_{x}=\sigma_{0}$, $\hat{M}_{y}=\sigma_{x}$, $\hat{C}_{3}=e^{\frac{2i\pi}{3}\tau_{z}\sigma_{z}}$, and $\hat{C}_{3}^{2}=e^{-\frac{i 2\pi}{3}\tau_{z}\sigma_{z}}$. One can easily confirm $\mathcal{T}$ and  $\mathcal{R} =[C_{2},C_{3},C^{2}_{3},M_{x}]$ symmetries  are present since  $\hat{H}_{\boldsymbol{K}^{\prime}}\left(-\boldsymbol{k}\right)=\mathcal{\hat{T}}
\hat{H}_{\boldsymbol{K}}\left(
\boldsymbol{k}\right)\mathcal{\hat{T}}^{-1}$ and $\hat{H}_{\mathcal{R}\boldsymbol{K}}\left(\mathcal{R}\boldsymbol{k}\right)=\mathcal{\hat{R}}
\hat{H}_{\boldsymbol{K}}\left(
\boldsymbol{k}\right)\mathcal{\hat{R}}^{-1}$ are satisfied, 
whereas the $M_{y}$ symmetries is absent due to
$\hat{H}_{\boldsymbol{K}}(k_{x},k_{y})\neq\hat{M}_{y}^{-1}\hat{H}_{\boldsymbol{K}^{\prime}}
(k_{x},-k_{y})\hat{M}_{y}$. 

For simplicity, we only focus on $n$-doped systems.
The corresponding energy dispersion $\varepsilon_{\tau_{v}}\left(\boldsymbol{k}\right)$  and BCP elements $\tilde{\mathcal{G}}_{\tau_v,ab}(\boldsymbol{k})$ [Eq.~\eqref{Gmaj}] for the conduction band of the monolayer graphene are found to be, respectively
\begin{equation}
\begin{aligned}
\varepsilon_{\tau_{v}}\left(\boldsymbol{k}\right)&=\hbar v_{F}k\\
\tilde{\mathcal{G}}_{\tau_{v},ab}(\boldsymbol{k})&=\tilde{\mathcal{G}}_{ab}
(\boldsymbol{k})=\frac{e}{4\hbar v_{F}k^{5}}
\left[
\begin{array}{cc}
    k_{y}^{2}       &-k_{x}k_{y} \\
    -k_{x}k_{y}     &k_{x}^{2}
   \end{array}
   \right],
\end{aligned}
\label{energlable}
\end{equation}
showing that the both $\varepsilon_{\tau_{v}}\left(\boldsymbol{k}\right)$ and $\tilde{\mathcal{G}}_{\tau_{v},ab}(\boldsymbol{k})$ are independent of the valley index $\tau_{v}$.  Consequently, the $\xi^\text{BCP}_{abcd}$ [Eqs.~\eqref{c-chi} and  \eqref{coeffu}] at two valleys can have the same sign and contributes additively to NANE. 
The BCP elements [$\tilde{\mathcal{G}}_{xx}$, $\tilde{\mathcal{G}}_{xy}$, $\tilde{\mathcal{G}}_{yx}$ and $\tilde{\mathcal{G}}_{yy}$] for the monolayer graphene reaches their maximum around the Dirac cone [Fig.~\ref{fig-BCP}].  

\begin{figure}[htbp]
\centering
\flushleft
\includegraphics[width=1.0\linewidth,clip]{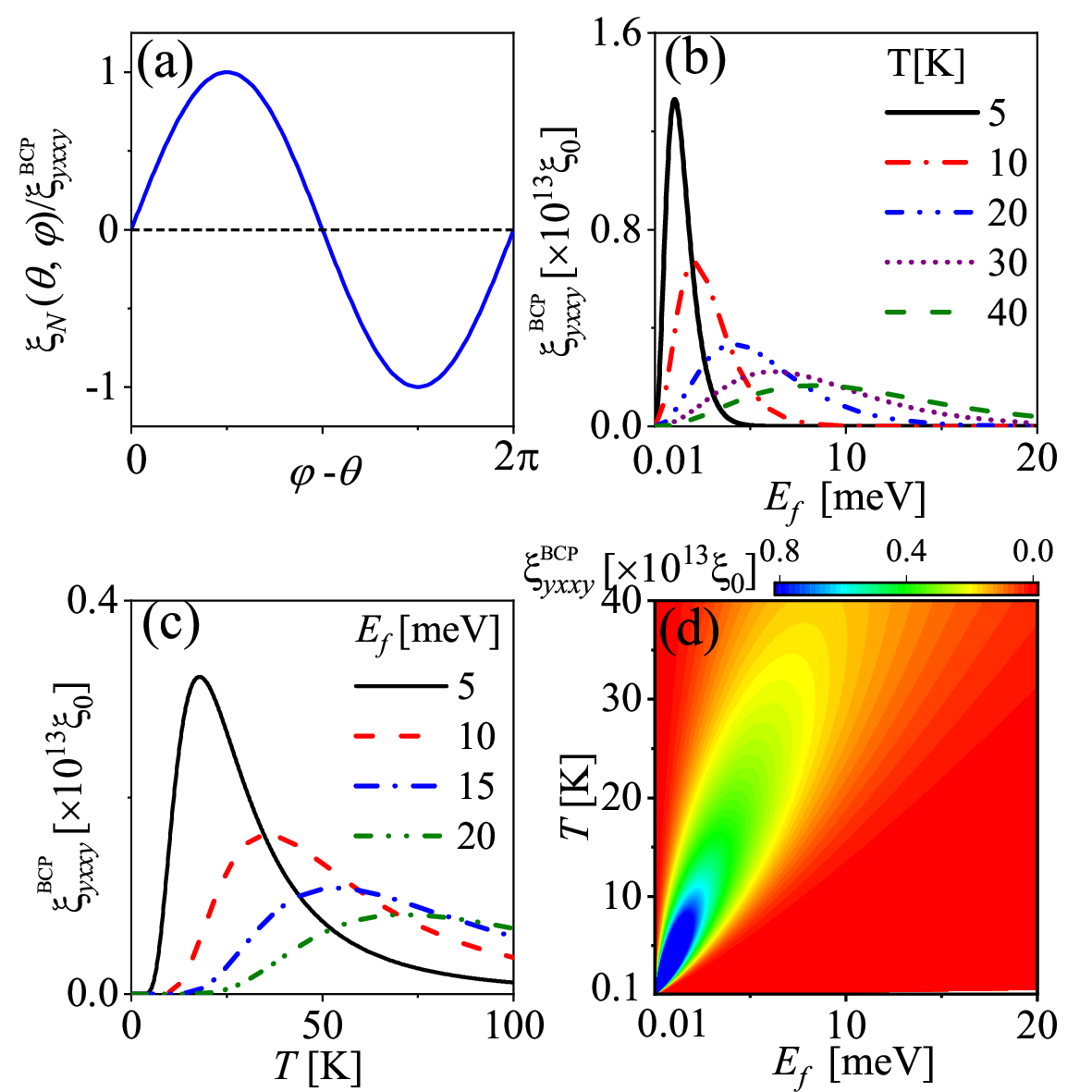}
\caption{(a) The dependence of $\xi_{\text{N}}(\theta,\varphi)$ on the relative orientation (i.e., $\varphi-\theta$) of dc electric field $\boldsymbol{E}^{\mathrm{dc}}$ with respective to temperature gradient.
 (b) $\xi^\text{BCP}_{yxxy}$ versus the Fermi energy $E_{f}$ for different temperature $T$.
 (c) $\xi^\text{BCP}_{yxxy}$ vs $T$ for different $E_{f}$ .
 (d) $\xi^\text{BCP}_{yxxy}$ as a function of $E_{f}$ and $T$.
  The relative angle $\varphi-\theta=\pi/2$ is fixed in (b)-(d).
  The typical scale $\xi_{0}$ is defined as $\xi_{0}=2\tau e^{2}k_{B}^{2}\alpha^{2}/[(1+i\omega\tau)\hbar^{2}t_{0}]$.
   }
\label{fig-2}
\end{figure}

The $C_{6v}$ symmetry of monolayer graphene guarantees that the angle-dependent scalar coefficient of  electric-field-induced NANC $\xi_\text{N}(\theta,\varphi)$ [Eq.~{\eqref{angel}}] only depends on the relative orientation of the applied $\boldsymbol{E}^\text{dc}$ to the temperature gradient [Fig.~\ref{figure-sch}]. Besides, the presence of both $\mathcal{T}$ and $\mathcal{P}$ symmetries forces $\xi^\text{NDF-AV}_{abcd}=0$ (see details in Sec.~\ref{SA-AD}). Hence, the $\xi_\text{N}(\theta,\varphi)$ in the monolayer graphene is reduced to be
\begin{equation}
\xi_\text{N}(\theta,\varphi)=\xi^\text{BCP}_{yxxy}\sin(\varphi-\theta),
\label{D-1}
\end{equation}
showing that the NANE comes exclusively from the contribution from BCP [$\xi^\text{BCP}_{yxxy}$] and exhibits a sine function dependence on the relative angle [$\varphi-\theta$] between the temperature gradient and $\boldsymbol{E}^{\text{dc}}$ [Fig.~\ref{fig-2}(a)]. When applying $\boldsymbol{E}^{\text{dc}}$ vertical to the temperature gradient (i.e., $\varphi-\theta=\pi/2,3\pi/2$), the quantity $|\xi_\text{N}(\theta,\varphi)|$ will reach its maximum, resulting in an extreme value of NANE in monolayer graphene. Conversely, when $\boldsymbol{E}^{\text{dc}}$ is parallel or antiparallel to the temperature gradient, the NANE disappears [$\xi_\text{N}(\theta,\varphi)=0$]. A typical scale $\xi_{0}=2\tau e^{2}k_{B}^{2}\alpha^{2}/[(1+i\omega\tau)\hbar^{2}t_{0}]$ (where the factor $2$ accounts for the spin degeneracy) is introduced to characterize the electric-field-induced NANC $\xi^\text{BCP}_{yxxy}$.

Fig.~\ref{fig-2}(b) illustrates the dependence of coefficient $\xi^\text{BCP}_{yxxy}$ on the Fermi energy $E_{f}$ for different temperature $T$.
A peak feature of $\xi^\text{BCP}_{yxxy}$ is observed. The emergence of this peak can be qualitatively attributed to the interplay between the rapid reduction of BCP and the linearly increasing density of states as the Fermi level moves away from the Dirac point (i.e. as $|E_{f}|$ increases).  As temperature $T$ decreases, the peak value increases and its position shifts towards lower energy levels [Figs.~\ref{fig-2}(c) and \ref{fig-2}(d)]. Hence, one might expect the larger values of $\xi^\text{BCP}_{yxxy}$ to be achieved at extremely low temperatures and extremely low Fermi energy levels (i.e., extremely low $n$ doping) [Fig.~\ref{fig-2}(d)]. {\color{blue}{The maximum of the color scale in \ref{fig-2}(d) has been artificially set to a low value (0.8) in order to clearly display the values in the high-Fermi-energy and high-temperature regime}}. However, our theory does not apply at extremely low Fermi levels because the underlying Boltzmann theory's criterion \cite{Chester-1963,Pottier-2010} $\hbar/\tau \ll E_{f}$ breaks down as $E_{f}\rightarrow 0$ close to Dirac point, and the experimentally unavoidable electron-hole puddles near the Dirac cone introduce spatial inhomogeneity that violates the homogeneity assumption required for the relaxation-time-approximation\cite{Adam-2007,Deal-2010}.
 In this regime, a quantum kinetic approach\cite{H. Varshney2023,Sekine-2020,Pankaj-2020} would be necessary, which is beyond the scope of the present work.

Nevertheless, considering the currently attainable high-quality monolayer graphene with inhomogeneity \cite{Mayorov-2012,Sarma-2013}$\delta n\sim10^{8}~\,\mathrm{cm}^{-2}$, impurity concentration $ n_\text{imp}\sim 3\times10^{9}\, \mathrm{cm}^{-2}$, and mobilities $\mu=0.5\sim2.5\times 10^{6} \,\mathrm{cm}^{2}\mathrm{V}^{-1}s^{-1}$ at carrier concentrations $n\sim1\times10^{10}\, \mathrm{cm}^{-2}$, our theory is applicable when $E_{f}>10\, \mathrm{meV}$. That's because (i) the low inhomogeneity enables the experiment access to the Dirac point down to $E^{*}_{f}=\hbar v_{F}\sqrt{\pi\delta n}\sim1.1 \,\mathrm{meV}$, implying electron-hole puddle effects are negligible for $E_{f}>E^{*}_{f}$, and (ii) the mean free path $l=(\hbar/e)\mu(\pi n)^{1/2}$ can reach $2.5\, \mathrm{\mu m}$, yielding $\tau=l/v_{F}\approx 2.5\, \mathrm{ps}$ and a relaxation length $\hbar/\tau\approx 0.264\,\mathrm{meV}$; hence, the criterion $\hbar/\tau\ll E_{f}$ is approximately satisfied for $E_{f}>10\, \mathrm{meV}$ ($\sim38\hbar/\tau)$. In practice, our Boltzmann-based theory can be considered applicable for carrier densities $n>n_{imp}\sim3\times10^9 \,\mathrm{cm}^{-2}$  [see appendix \ref{DF-T} for details], which corresponds to $E_{f}>6\,\mathrm{meV}$. If $n_{imp}$ can be made very low in further experiment, then our Boltzmann-based theory would apply down to very low density (or Fermi level) as long as interference and interaction effects are negligible\cite{Sarma-2013}.

 When applying $\boldsymbol{E}^\text{dc}$ vertical to $\boldsymbol{\nabla} {T}^{\omega}$ (i.e., $\varphi-\theta=\pi/2,3\pi/2$), the second-harmonic nonlinear Nernst current in monolayer graphene is found to be $j^{\mathrm{nl},2\omega}_{\mathrm{N}}=\xi^\text{BCP}_{yxxy}(\tilde{\partial{T}})^{2}E$.
 which corresponds to a second-harmonic Nernst voltage  $V ^{2\omega}_{\mathrm{N}}=W j^{\mathrm{nl},2\omega}_{\mathrm{N}}/\sigma$ (see the details in Appendix \ref{app-voltage}) in the open-circuit case with $W$ being the width of the sample. Experimentally, the amplitude of the second-harmonic transverse voltage $\text{Re}\{V ^{2\omega}_{N}\}$ is detected by lock-in amplification at frequency $2\omega$.
  The conductivity $\sigma=n_{c}\mu e$ can be estimated as $\tau E_f e^{2}/\hbar^{2}\pi$, since the mobility $\mu$ is expressed as  $\tau e/m$, and meanwhile, for the  monolayer graphene, we have the effective mass $m=\hbar\sqrt{\pi n_{c}}/v_{F}$ and the carrier density $n_{c}=E_{f}^{2}/(v_{F}^{2}\hbar^{2}\pi)$. \cite{Castro-Neto2009}
To numerically estimate the signal of the second-harmonic voltage drop originated from the electric-field-induced NANE in monolayer graphene, we take $\xi^\text{BCP}_{yxxy}=0.1611\times10^{13}\xi_{0}$
for $T=35~\mathrm{K}$ and $E_{f}=10~\mathrm{meV}$. In the experiment, the real part of amplitude of the alternating temperature gradient $\text{Re}\{\tilde{\partial{T}}\}$ can already reach 10~$\mathrm{mK}\mathrm{\mu m}^{-1}$,\cite{Liu-2025} {\color{blue}{and the dc electric field can reach $E^{\mathrm{dc}}=0.05~\mathrm{V/\mu m}$\cite{J.Yang2025}. Here, we adopt $E^{\mathrm{dc}}=4\times10^{-4}~\mathrm{V/\mu m}$ since the weak-field condition $\eta=eE^{\mathrm{dc}}l/E_f\ll1$ is required for the first-order expansion of equilibrium Fermi distribution in electric field [see appendix \ref{DF-T} for details]. For the mean free path $l\approx2.5~\mathrm{\mu m}$ and $E_f=10~\mathrm{meV}$, this choice yields $\eta\approx0.1$, which is marginally sufficient for the first-order weak-field treatment.}} Therefore, in the low frequency limit $\omega\tau<<1$, the amplitude of the second-harmonic voltage ($\mathrm{Re}\{V^{2\omega}\}$) becomes frequency-independent since its frequency component $\text{Re}\{{1}/(1+i\omega\tau)\}\approx1$ and can reach $1.2~\mathrm{\mu V}$ for $W=10~\mathrm{\mu m}$, which is measurable {\color{blue}{\cite{Uchida}}}.


\section{CONCLUSION} \label{fou}
In summary, we have investigated the nonlinear Nernst effect as a second-order response to a alternating temperature gradient $\boldsymbol{\nabla} {T}^{\omega}$ under the control of the dc electric field $\boldsymbol{E}^\text{dc}$. It's found that a nonzero electric-field-induced NANE, linearly dependent on the relaxation time, can emerge owing to electric-field-corrected Berry curvature or the anomalous-velocity-modified NDF in $\mathcal{T}$-symmetric 2D materials with high crystal symmetries, where the intrinsic and ${\mathbf{\Lambda}}^{T}$-induced  inherent  NANE is prohibited. Specially, we identify that when $\boldsymbol{E}^\text{dc}\perp \boldsymbol{\nabla} {T}^{\omega}$, the electric-field-induced NANE is permissible across all the crystallographic symmetries applicable to 2D systems. The electric-field-induced NANE shows strong angular dependence on the orientation of $\boldsymbol{E}^\text{dc}$ with respect to $\boldsymbol{\nabla} {T}^{\omega}$, offering a route to control its signal. The behaviors of the NANE in the monolayer graphene have been investigated. Due to the simultaneous presence of $\mathcal{T}$ and $\mathcal{P}$ symmetry, both the  $\boldsymbol{\Lambda}^{T}$-induced inherent NANE and the electric-field-induced NANE originating from anomalous-velocity-modified-NDF contribution vanish, resulting in the electric-field-induced NANE from BCP determining the second-harmonic Nernst voltage in pristine monolayer graphene.  While we focus on the 2D systems, the electric-field-induced NANE from BCP proposed here might also be expected in 3D centrosymmetric materials with time-reversal symmetry, such as the Dirac-semimetal $\mathrm{Cd_{3}As_{2}}$, where electric-field-induced BCD has been experimentally confirmed\cite{Zhao-2023}.
\section{acknowledgements}
This work is supported by the National Natural Science Foundation of China (Grant No. 12574073), and the National Science Foundation of Hunan, China (Grant No. 2023JJ30118)

\appendix

\setcounter{equation}{0}
\setcounter{figure}{0}
\setcounter{table}{0}
\makeatletter
\renewcommand{\theequation}{A\arabic{equation}}
\renewcommand{\thefigure}{A\arabic{figure}}
\renewcommand{\thetable}{A\arabic{table}}

\bigskip
\bigskip

\noindent

\section{The distribution function $f(\boldsymbol{k})$ in the presence of temperature gradient and electric field }
\label{DF-T}
Within the relaxation time approximation, the Boltzmann equation for the electron distribution $f(\boldsymbol{r},\boldsymbol{k},t)$, which is a function of position $\boldsymbol{r}$, momentum $\boldsymbol{k}$, and time $t${\color{blue}{,}} is
\begin{equation}
\begin{aligned}
\{\partial_{t}+\dot{\boldsymbol{r}}\cdot\nabla_{\boldsymbol{r}}+\dot{\boldsymbol{k}}\cdot\nabla_{\boldsymbol{k}}\}f(\boldsymbol{r},\boldsymbol{k},t)
\approx-\frac{f(\boldsymbol{r},\boldsymbol{k},t)-f_{0}(\boldsymbol{k})}{\tau},
\label{Ap-A-1}
\end{aligned}
\end{equation}
where $f_{0}=1/[\exp[(\varepsilon_{\boldsymbol{k}}-E_{f})/k_{B}T]+1]$ indicates the equilibrium
 Fermi distribution in absence of the external field, and $\tau$ is the relaxation time. Formally, the Boltzmann theory is valid for $\hbar/\tau \ll E_{f}$, which can be rewritten equivalently in the form of the Ioffe-Regel criterion: $k_{F}l\gg1$ with $k_{F}$ representing the Fermi wave vector and $l$ indicating the mean free path \cite{Chester-1963,Pottier-2010}. In practice, however, the theory can still be considered reliable as long as the carrier density $n$ exceeds the impurity concentration $n_{imp}$, i.e., $n>n_{imp}$.\cite{Sarma-2013,Hwang-2007} Besides, for the relaxation-time approximation in the Boltzmann equation, the system is required to be spatially homogeneous\cite{Adam-2007,Deal-2010}.

  In absence of magnetic field, the semiclassical equation of the motion is
\begin{equation}
\begin{aligned}
&\dot{\boldsymbol{r}}=\frac{1}{\hbar}\frac{\partial \varepsilon_{\boldsymbol{k}}}
{\partial\boldsymbol{k}}-\dot{\boldsymbol{k}}\times \mathbf{\tilde{\Omega}}\left(\boldsymbol{k}\right),\\
&\dot{\boldsymbol{k}}=-\frac{e}{\hbar}\boldsymbol{E},
\label{Ap-A-2}
\end{aligned}
\end{equation}
where $\tilde{\boldsymbol{\Omega}}(\boldsymbol{k})=\boldsymbol{\Omega}(\boldsymbol{k})+\boldsymbol{\Omega}
^{E}(\boldsymbol{k})$ represents the corrected Berry curvature by electric field with  $\boldsymbol{\Omega}(\boldsymbol{k})$ and $\boldsymbol{\Omega}^{E}(\boldsymbol{k})\propto \boldsymbol{E}$ representing the inherent and electric-field-corrected Berry curvature, respectively. We considers an \textit{alternating} temperature gradient $\boldsymbol{\nabla} {T}^{\omega}(t)=\text{Re}\{ \tilde{\boldsymbol{\nabla} T} e^{i\omega t}\}$ (where complex amplitude vector
$\tilde{\boldsymbol{\nabla} T}$ is independent of the time), which can be experimentally  realized by passing an alternating current $I^{\omega^{\prime}}_{h}=I_{0}\sin(\omega^{\prime}t+\phi_{0})$ ($\omega^{\prime}=\omega/2$, $\phi_{0}$ indicates the initial phase) through a  microheater electrode\cite{Liu-2025}. The corresponding complex amplitude vector is given by
 $\tilde{\boldsymbol{\nabla} T}=\hat{n}QI_{0}^{2}e^{i\pi+{\color{blue}{2}}i\phi_{0}}/2$ with $Q$ representing proportionality coefficient and $\hat{n}$ indicating the unit vector along the temperature gradient. The local distribution function $f(\boldsymbol{r},\boldsymbol{k},t)$ fixed itself by the temperature at $\boldsymbol{r}$, can be then determined as

\begin{equation}
\begin{aligned}
\nabla_{\boldsymbol{r}} f=\nabla_{\boldsymbol{r}}T(t) \frac{\partial f}{\partial T}=\text{Re}\{\tilde{\boldsymbol{\nabla}T} e^{i\omega t}\}\frac{\partial f}{\partial T}.
\label{Ap-A-3}
\end{aligned}
\end{equation}
Combining Eqs.~(\ref{Ap-A-1}) (\ref{Ap-A-2}) and Eq. (\ref{Ap-A-3}), we have
\begin{equation}
\begin{aligned}
\frac{f_{0}-f}{\tau}
=&\left[\left(\frac{1}{\hbar}\frac{\partial \varepsilon_{\boldsymbol{k}}}
{\partial\boldsymbol{k}}+\frac{e}{\hbar}(\boldsymbol{E}\times\mathbf{\tilde{\Omega}})
\right)\cdot\text{Re}\{\tilde{\boldsymbol{\nabla}T}e^{i\omega t}\}\frac{\partial }{\partial T} \right.  \\
& \left.
    -\frac{e}{\hbar}\boldsymbol{E}\cdot\nabla_{\boldsymbol{k}}+\partial_{t}\right]f.
\label{Ap-A-4}
\end{aligned}
\end{equation}
$f$ can be expressed as a perturbation expansion:
\begin{equation}
\begin{aligned}
f=f_{0}+\sum_{n,m}\delta f^{(n,m)},
\end{aligned}
\label{Ap-A-5}
\end{equation}
where $\delta f^{(n,m)}$ represents the component of nonequilibrium distribution corresponding to the $n$-th order in electric field $E$ and $m$-th order in the amplitude of temperature gradient $\text{Re}\{\tilde{\boldsymbol{\nabla}T}\}$. Taking Eq.~(\ref{Ap-A-5}) into Eq.~(\ref{Ap-A-4}) and comparing the expansion orders, one can obtain a recursive equation for $\delta f^{(n,m)}$ as
\begin{widetext}
\begin{equation}
\begin{aligned}
-\frac{\delta f^{(n,m)}}{\tau}=&\partial_{t}f^{(n,m)}-\frac{e}{\hbar}\boldsymbol{E}\cdot\nabla_{\boldsymbol{k}} \delta f^{(n-1,m)}+\sum_{l}(\frac{1}{\hbar}\frac{\partial \varepsilon_{\boldsymbol{k}}}
{\partial\boldsymbol{k}})^{(l)}
\cdot \frac{1}{2}\left[\tilde{\boldsymbol{\nabla}T} e^{i\omega t}+ \left(\tilde{\boldsymbol{\nabla}T}\right)^{*} e^{-i\omega t}\right]\frac{\partial \delta f^{(n-l,m-1)}}{\partial T}
\\
& +\sum_{l}\frac{e}{\hbar}(\boldsymbol{E}\times \mathbf{\tilde{\Omega}})^{(l)}
\cdot  \frac{1}{2}\left[\tilde{\boldsymbol{\nabla}T} e^{i\omega t}+ \left(\tilde{\boldsymbol{\nabla}T}\right)^{*} e^{-i\omega t}\right]\frac{\partial  \delta f^{(n-l,m-1)}}{\partial T}.
\label{Ap-A-6}
\end{aligned}
\end{equation}
\end{widetext}

We are interested in the response to second order in the temperature gradient and the first order in the electric field. Therefore, we will only calculate the distribution up to the second order in the temperature gradient and the first order in the electric field, namely $f\approx\mathrm{Re}\{f_{0}+\delta f^{(1,0)}+\delta f^{(0,1)}+\delta f^{(1,1)}+\delta f^{(0,2)}\}$. {\color{blue}{This expansion in electric field  is valid only in the week-field regime, namely the field-induced change in electron energy ($e\mathbf{E}^\text{dc}\cdot \mathbf{r}$) must be very weak compared to Fermi energy $E_{F}$. The spatial coordinate can be written as $\mathbf{r}=\mathbf{v}_{b}\tau$ with $\mathbf{v}_{b}$ representing the band velocity \cite{Aniru-2019,Kapri-2025}. For a linear Dirac band, one has $|\mathbf{v}_{b}|=v_{F}$. Hence, the weak-field condition for the expansion approach requires the dimensionless parameter $\eta=e E^\text{dc} l/E_{f}<<1$ with $l=\tau v_{F}$ indicating the mean free path.  }}

 By iteration, these leading orders [ $\delta f^{(1,0)}$, $\delta f^{(0,1)}$, $\delta f^{(1,1)}$ and $\delta f^{(0,2)}$] are determined as, respectively
\begin{equation}
\begin{aligned}
&\delta f^{(1,0)}=\frac{e\tau}{\hbar}\boldsymbol{E}\cdot\nabla_{\boldsymbol{k}}f_{0},~~~~~
\delta f^{(0,1)}=\delta f^{(0,1)}_{\omega}e^{i\omega t},\\
&\delta f^{(1,1)}=\delta f^{(1,1)}_{\omega}e^{i\omega t},~~~~\delta f^{(0,2)}=\delta f^{(0,2)}_{0}+\delta f^{(0,2)}_{2\omega}e^{2i\omega t},\\
\end{aligned}
\label{app-a-f}
\end{equation}
with
\begin{equation}
\begin{aligned}
\delta f^{(0,1)}_{\omega}&=-\frac{\tau \boldsymbol{v}\cdot\tilde{\boldsymbol{\nabla}T}}{(1+i\omega\tau)}\frac{\partial f_{0}}{\partial T},\\
\delta f^{(1,1)}_{\omega}&=\frac{\tau}{(1+i\omega\tau)}\left[\frac{e}{\hbar}\boldsymbol{E}
\cdot\nabla_{\boldsymbol{k}}\delta f^{(0,1)}_{\omega}-\boldsymbol{v}\cdot\tilde{\boldsymbol{\nabla}T}\frac{\partial \delta f^{(1,0)}}{\partial T} \right.\\
   &\left. -\left(\frac{e}{\hbar}\boldsymbol{E}\times\mathbf{\Omega}\right)\tilde{\cdot
                 \boldsymbol{\nabla}T}\frac{\partial f_{0}}{\partial T}\right],\\
\delta f^{(0,2)}_{0}&=-\frac{\tau\boldsymbol{v}\cdot(\tilde{\boldsymbol{\nabla}T})^{*}}{2}\frac{\partial \delta f^{(0,1)}_{\omega}}{\partial T},\\
\delta f^{(0,2)}_{2\omega}&=-\frac{\tau\boldsymbol{v}\cdot\tilde{\boldsymbol{\nabla}T}}{2(1+2i\omega\tau)}\frac{\partial \delta f^{(0,1)}_{\omega}}{\partial T},
\end{aligned}
\label{Ap-A-7}
\end{equation}
where the equality $\partial \varepsilon_{\boldsymbol{k}}/\partial \boldsymbol{k}=\hbar \boldsymbol{v}$ has been used for simplicity. One might notice the following relation
\begin{equation}
\begin{aligned}
\frac{\partial f_{0}}{\partial T}=-\frac{\left(\varepsilon_{\boldsymbol{k}}-E_{f}\right)}{T}\frac{\partial f_{0}}{\partial \varepsilon_{\boldsymbol{k}}}.
\end{aligned}
\label{Ap-A-8}
\end{equation}
Combining Eq.~(\ref{Ap-A-8}) with equality $\partial \varepsilon_{\boldsymbol{k}}/\partial \boldsymbol{k}=\hbar \boldsymbol{v}$, we can further transform $\partial f_{0}/\partial T$ into $\partial f_{0}/\partial \boldsymbol{k}$ through the differential treatment, namely
\begin{equation}
\frac{\partial f_{0}}{\partial \boldsymbol{k}}=\frac{\partial f_{0}}{\partial \varepsilon_{\boldsymbol{k}}} \frac{\partial \varepsilon_{\boldsymbol{k}}}{\partial \boldsymbol{k}}=-\frac{\hbar \boldsymbol{v}T}{\left(\varepsilon_{\boldsymbol{k}}-E_{f}\right)}\frac{\partial f_{0}}{\partial T},
\label{Ap-A-9}
\end{equation}
giving the following identities
\begin{equation}
\begin{aligned}
\frac{\partial f_{0}}{\partial T} v_{a}&=-\frac{\varepsilon_{\boldsymbol{k}}-E_{f}}{\hbar T}\frac{\partial f_{0}}{\partial k_{a}},\\
\frac{\partial^{2}f_{0}}{\partial T^{2}}v_{a}v_{b}
&=2\frac{\varepsilon_{\boldsymbol{k}}-E_{f}}{\hbar T^{2}}\frac{\partial f_{0}}{\partial k_{a}}v_{b}+\left(\frac{\varepsilon_{\boldsymbol{k}}-E_{f}}{\hbar T}\right)^{2}\frac{\partial^{2}f_{0}}{\partial k_{a}\partial k_{b}}.
\end{aligned}
\label{Ap-A-10}
\end{equation}
Using these two identities, the coefficients [$\delta f^{(0,1)}_{\omega}$, $\delta f^{(1,1)}_{\omega}$, $\delta f^{(0,2)}_{0}$ and $\delta f^{(0,2)}_{2\omega}$] in Eq.~(\ref{Ap-A-7}) can be further simplified as
\begin{align}
\delta f^{\left(0,1\right)}_{\omega}&=\frac{\tau\left(\varepsilon_{\boldsymbol{k}}-E_{f}\right)}{\left(1+
i\omega\tau\right)T\hbar}\frac{\partial f_{0}}{\partial k_{a}}\tilde{\partial_{a}T},\notag\\
\delta f^{\left(1,1\right)}_{\omega}&=\left\{\frac{e\tau^{2}\left(2+i\omega\tau\right)}{\left(1+
i\omega\tau\right)^{2}\hbar^{2}T}\left[(\varepsilon_{\boldsymbol{k}}-E_{f})\frac{\partial^{2}f_{0}}
{\partial k_{a}\partial k_{b}}+\hbar v_{b}\frac{\partial f_{0}}{\partial k_{a}} \right]\right.\notag\\
 &\left. +\epsilon_{abc}\frac{e\tau}{\left(1+i\omega\tau\right)\hbar}\Omega_{c}\frac{
 \left(\varepsilon_{\boldsymbol{k}}-E_{f}\right)}{T}\frac{\partial f_{0}}{\partial \varepsilon_{\boldsymbol{k}}} \right\}  \tilde{\partial_{a}T}E_{b},\notag\\
\delta f^{(0,2)}_{0}&=\frac{\tau^{2}}{2(1+i\omega\tau)}
    \left[\frac{2\left(\varepsilon_{\boldsymbol{k}}-E_{f}\right)}{\hbar T^{2}}\frac{\partial f_{0}}{\partial k_{a}}v_{b}+\frac{\left(\varepsilon_{\boldsymbol{k}}-E_{f}\right)^{2}}{\hbar^{2}T^{2}} \right.\notag\\
  &\left. \times\frac{\partial^{2}f_{0}}{\partial k_{a}\partial k_{b}}\right]
       \left(\tilde{\partial_{a}T}\right)^{*}\tilde{\partial_{b}T},\notag\\
\delta f^{(0,2)}_{2\omega}&=\frac{\tau^{2}}{2(1+i\omega\tau)(1+2i\omega\tau)}
           \left[\frac{2\left(\varepsilon_{\boldsymbol{k}}-E_{f}\right)}{\hbar T^{2}}\frac{\partial f_{0}}{\partial k_{a}}v_{b} \right.\notag\\
 & \left.+\frac{\left(\varepsilon_{\boldsymbol{k}}-E_{f}\right)^{2}}{\hbar^{2} T^{2}}\frac{\partial^{2}f_{0}}{\partial k_{a}\partial k_{b}}\right]\tilde{\partial_{a}T}\tilde{\partial_{b}T}.
\label{Ap-A-11}
\end{align}

\setcounter{equation}{0}
\setcounter{figure}{0}
\setcounter{table}{0}
\makeatletter
\renewcommand{\theequation}{B\arabic{equation}}
\renewcommand{\thefigure}{B\arabic{figure}}
\renewcommand{\thetable}{B\arabic{table}}

\bigskip
\bigskip

\noindent
\noindent
\section{The electric-field-induced nonlinear anomalous Nernst effect in $\mathcal{T}$-symmetric systems}
\label{ESNAE}
Accompanying Eq.~(\ref{curre0})  with Eqs.~(\ref{app-a-f}) and (\ref{Ap-A-11}), the thermally driven nonlinear current density  ${j}^{\mathrm{nl}}_{a}$ in $a$ direction, as a second-order response to the alternating temperature gradient ${\boldsymbol{\nabla} T^{\omega}}$ in the presence of a dc electric field $\boldsymbol{E}^\text{dc}$, can be expressed as
\begin{equation}
{j}^{\mathrm{nl}}_{a}=\mathrm{Re}\{{j}^{\mathrm{nl},0}_{a}+
{j}^{\mathrm{nl},2\omega}_{a}e^{2i\omega t}\},
\label{App-C-1}
\end{equation}
where the rectified current $j_{a}^{\mathrm{nl},0}$ and the second-harmonic term $j_{a}^{\mathrm{nl},2\omega}$ are determined as, respectively,
\begin{equation}
\begin{aligned}
j_{a}^{\mathrm{nl},0}=&-e\int[d\boldsymbol{k}]\dot{r}_{a}\delta f^{(0,2)}_{0}-\epsilon_{abl}\frac{e}{2\hbar T}\int[d\boldsymbol{k}]\tilde{\Omega}_{l}(\boldsymbol{k})
 \\
&\times(\varepsilon_{\boldsymbol{k}}-E_{f}) \left[\delta f^{(0,1)}_{\omega}+\delta f^{(1,1)}_{\omega}\right]
(\tilde{\partial_{b}T})^{*}, \\
j_{a}^{\mathrm{nl},2\omega}=&-e\int[d\boldsymbol{k}]\dot{r}_{a}
  \delta f^{(0,2)}_{2\omega}  -\epsilon_{abl}\frac{e}{2\hbar T}\int[d\boldsymbol{k}]\tilde{\Omega}_{l}(\boldsymbol{k})\\
  &\times(\varepsilon_{\boldsymbol{k}}-E_{f})\left[\delta f^{(0,1)}_{\omega}+\delta f^{(1,1)}_{\omega}\right]\tilde{\partial_{b}T},\\
\end{aligned}
\label{App-C-2}
\end{equation}
where $\epsilon_{abl}$ is the Levi-Civita symbol, the indices ($a,~b,~l$) represent the cartesian components and the repeated indices are summed over.
According to Eq.~(\ref{Ap-A-11}), one can easily observe that both $\text{Re}\{\delta f^{(0,1)}_{\omega}\}$ (linear to $\tau$) and $\text{Re}\{\delta f^{(1,1)}_{\omega}\}$ (one component $\propto\tau$ while others $\propto\tau^{{2}}$) are linearly dependent of the amplitude of temperature gradient $\text{Re}\{\tilde{\partial_{d}T}\}$, while $\text{Re}\{\delta f^{(0,2)}_{2\omega}\}$ and $\text{Re}\{\delta f^{(0,2)}_{0}\}$  (quadratic to $\tau$) quadratically depends on $\text{Re}\{\tilde{\partial_{d}T}\}$.
Under $\mathcal{T}$ symmetry, the current ${j}^{\mathrm{nl}}_{a}$ and $\tau$ change sign but temperature gradient keeps unchanged, restricting the $\tau$-even dependent component of the current vanishing \cite {WU-2025,Gao-2019}. Therefore, only the terms related to $\delta f^{(0,1)}_{\omega}$ and $\tau$-odd component of $\delta f^{(1,1)}_{\omega}$ [Eq.~(\ref{Ap-A-11})]
in Eq.~(\ref{App-C-2}) contribute to thermally driven nonlinear current ${j}^{\mathrm{nl}}_{a}$  in $\mathcal{T}$- symmetric systems, which is our focus in this work. One can also confirm that these nonzero nonlinear current are vertical to temperature gradient, namely the Nernst current, since the first term ($\propto\delta f^{(0,2)}_{0}$) in Eq.~\eqref{App-C-2} has no contribution in $\mathcal{T}$-symmetric systems while the nonlinear current from the second terms are actually perpendicular to the temperature gradient. Consequently, after a tedious derivation, the nonzero components $j_{a}^{\mathrm{nl},0}$ and $j_{a}^{\mathrm{nl},2\omega}$ of the nonlinear Nernst current, to first-order approximation of the electric field, in $\mathcal{T}$-symmetric systems are found to be
\begin{equation}
\begin{aligned}
j_{a}^{\mathrm{nl},0}&=\left(\chi_{abc}+\xi_{abcd}E_{d}\right)(\tilde{\partial_{b}
T})^{*}\tilde{\partial_{c}T},\\
j_{a}^{\mathrm{nl},2\omega}&=\left(\chi_{abc}+\xi_{abcd}E_{d}\right) \tilde{\partial_{b}T}\tilde{\partial_{c}T}
\end{aligned}
\label{App-C-JTO}
\end{equation}
with
\begin{equation}
\begin{aligned}
\chi_{abc}=&\frac{-\tau e\epsilon_{abl}}{2(1+i\omega\tau)\hbar^{2}} \int[d\boldsymbol{k}]{\Omega}_{l}\left(\boldsymbol{k}\right)
     \frac{(\varepsilon_{\boldsymbol{k}}-E_{f})^{2}}{T^{2}}\frac{\partial f_{0}}{\partial k_{c}},\\
\xi_{abcd}=&\frac{-\tau e\epsilon_{abl}}{2(1+i\omega\tau)\hbar^{2}}\left[
\epsilon_{lgv}\!\!\int[d\boldsymbol{k}]
\frac{\partial{\tilde{\mathcal{G}}_{vd}\left(\boldsymbol{k}\right)}}{\partial k_{g}}\frac{(\varepsilon_{\boldsymbol{k}}-E_{f})^{2}}{T^{2}}\frac{\partial f_{0}}{\partial k_{c}}
     \right.\\
&\left.
 +e\epsilon_{cdu}\int[d\boldsymbol{k}] \Omega_{l}(\boldsymbol{k})\Omega_{u}(\boldsymbol{k}) \frac{(\varepsilon_{\boldsymbol{k}}-E_{f})^{2}}{T^{2}}\frac{\partial f_{0}}{\partial \varepsilon_{\boldsymbol{k}}}
 \right].
\end{aligned}
\label{App-C-to}
\end{equation}
Obviously, both nonzero coefficients $\chi_{abc}$ and $\xi_{abcd}$ in $\mathcal{T}$-symmetric systems have been restricted to be odd functions of $\tau$, hinting  $\chi_{abc}$ and $\xi_{abcd}$ $\mathcal{T}$-odd. Additionally, one can also intuitively confirm this $\mathcal{T}$-odd character based on Eq.~\eqref{App-C-JTO}. Under $\mathcal{T}$ symmetry, only ${j}^{\mathrm{nl}}_{a}$ changes sign, while both temperature gradient and electric field keep unchanged, enforcing  $\chi_{abc}$ and $\xi_{abcd}$ $\mathcal{T}$-odd.

Equation~{\eqref{App-C-to}} shows the coefficient $\chi_{abc}$ is proportional to a pseudotensorial quantity $\Lambda^{T}_{lc}$, defined as
\begin{equation}
\Lambda^{T}_{lc}=-\int[d\boldsymbol{k}]{\Omega}_{l}\left(\boldsymbol{k}\right)
     \frac{(\varepsilon_{\boldsymbol{k}}-E_{f})^{2}}{T^{2}}\frac{\partial f_{0}}{\partial k_{c}},
\label{App-C-lc}
\end{equation}
which was firstly defined and analyzed by Yu \textit{et al}.\cite{X.-Q. Yu2019}, and play as the quantum origin to generate the inherent NANE. This novel pseudotensorial quantity
$\Lambda_{lc}^{T}$ stems from the Berry curvature near the Fermi surface, and features an extra factor $(\varepsilon_{\boldsymbol{k}}-E_{f})^{2}/T^{2}$ in integral comparing the BCD\cite{I. Sodemann2015}, $D_{lc}=-\int[d\boldsymbol{k}]{\Omega}_{l}\left(\boldsymbol{k}\right)\partial_{k_{c}} f_{0}$. The first term of the coefficient $\xi_{abcd}$ in Eq.~\eqref{App-C-to} comes from the BCP rooting in the electric-field-induced Berry curvature $\boldsymbol{\Omega}^{E}(\boldsymbol{k})$, and the second term actually comes from the anomalous velocity ($\boldsymbol{E}\times\mathbf{\Omega}$) corrected NDF. 
The coefficients  $\xi_{abcd}$ thus can be decomposed into two parts as
\begin{equation}
 \xi_{abcd}=\frac{\tau e\epsilon_{abl}}{2(1+i\omega\tau)\hbar^{2}}\left(\Upsilon_{lcd}
 +e\epsilon_{cdu}\Gamma_{lu}\right)
\label{app-c-chi}
\end{equation}
with
\begin{equation}
\begin{aligned}
\Upsilon_{lcd}&=-\epsilon_{lgv}\int[d\boldsymbol{k}]
\frac{\partial{\tilde{\mathcal{G}}_{vd}\left(\boldsymbol{k}\right)}}{\partial k_{g}}
     \frac{(\varepsilon_{\boldsymbol{k}}-E_{f})^{2}}{T^{2}}\frac{\partial f_{0}}{\partial k_{c}},\\
\Gamma_{lu}&=-
\int[d\boldsymbol{k}] \Omega_{l}(\boldsymbol{k})\Omega_{u}(\boldsymbol{k}) \frac{(\varepsilon_{\boldsymbol{k}}-E_{f})^{2}}{T^{2}}\frac{\partial f_{0}}{\partial \varepsilon_{\boldsymbol{k}}}.
\end{aligned}
\label{App-coeffu}
\end{equation}
In 2D crystals, the inherent and field-corrected Berry curvature transforms from a pseudovector to a pseudoscalar, and only the out-of-plane component [i.e. $\Omega_{l=\mu=z}(\boldsymbol{k})$ and $\Omega_{l=z}^{E}\left(\boldsymbol{k}\right)$] can be nonzero, enforcing the indices $[l,\mu]$ in Eqs.~(\ref{App-C-to})-(\ref{App-coeffu}) fixed to be $z$. Consequently, the pseudotensorial quantity $\Lambda^{T}_{cd}$ behaves as a pseudovector, while the third-order pseudotensorial quantity $\Upsilon_{lcd}$ (pseudotensorial $\Gamma_{lu}$) behave as a pseudotensorial (scalar) quantity contained in the 2D plane, respectively:
\begin{equation}
\begin{aligned}
\Lambda^{T}_{c}&=-\int[d\boldsymbol{k}]{\Omega}_{z}\left(\boldsymbol{k}\right)
     \frac{(\varepsilon_{\boldsymbol{k}}-E_{f})^{2}}{T^{2}}\frac{\partial f_{0}}{\partial k_{c}},\\
\Upsilon_{cd}&=\int[d\boldsymbol{k}]
\left[\frac{\partial{\tilde{\mathcal{G}}_{xd}\left(\boldsymbol{k}\right)}}{\partial k_{y}}
-\frac{\partial{\tilde{\mathcal{G}}_{yd}\left(\boldsymbol{k}\right)}}{\partial k_{x}}\right]
     \frac{(\varepsilon_{\boldsymbol{k}}-E_{f})^{2}}{T^{2}}\frac{\partial f_{0}}{\partial k_{c}},\\
\Gamma&=-
\int[d\boldsymbol{k}] \Omega^{2}_{z}(\boldsymbol{k}) \frac{(\varepsilon_{\boldsymbol{k}}-E_{f})^{2}}{T^{2}}\frac{\partial f_{0}}{\partial \varepsilon_{\boldsymbol{k}}}.
\end{aligned}
\label{App-coeffs}
\end{equation}
As a result, the rectified current $\boldsymbol{j}^{\mathrm{nl},0}$ and the second-harmonic term $\boldsymbol{j}^{\mathrm{nl},2\omega}$ for the 2D systems can be written in vector notation as, respectively,
\begin{equation}
\begin{aligned}
\boldsymbol{j}^{\mathrm{nl},0}&=
\frac{e\tau}{2(1+i\omega\tau)\hbar^{2}}\left(\hat{z}\times(\tilde{\boldsymbol{\nabla}T})^{*}\right)
\left(\tilde{\boldsymbol{\nabla}T}\cdot\mathbf{\Lambda}^{T}\right)\\
& +\frac{e\tau}{2(1+i\omega\tau)\hbar^{2}}\left(\hat{z}\times(\tilde{\boldsymbol{\nabla}T})^{*}\right)
\left(\tilde{\boldsymbol{\nabla}{T}}\cdot\overleftrightarrow{\boldsymbol{\Upsilon}}\right)\cdot \boldsymbol{E}\\
&+\frac{e^{2}\tau\Gamma}{(1+i\omega\tau)\hbar^{2}}\left(\hat{z}\times(\tilde{\boldsymbol{\nabla}T}
)^{*}\right)\left(\hat{z}\times\tilde{\boldsymbol{\nabla}{T}}
\right)\cdot\boldsymbol{E} ,
\end{aligned}
\end{equation}
and
\begin{equation}
\begin{aligned}
\boldsymbol{j}^{\mathrm{nl},2\omega}&=
\frac{e\tau}{2(1+i\omega\tau)\hbar^{2}}\left(\hat{z}\times\tilde{\boldsymbol{\nabla}T}\right)
\left(\tilde{\boldsymbol{\nabla}T}\cdot\mathbf{\Lambda}^{T}\right)\\
& +\frac{e\tau}{2(1+i\omega\tau)\hbar^{2}}\left(\hat{z}\times\tilde{\boldsymbol{\nabla}T}\right)
\left(\tilde{\boldsymbol{\nabla}{T}}\cdot\overleftrightarrow{\boldsymbol{\Upsilon}}\right)\cdot \boldsymbol{E}\\
&+\frac{e^{2}\tau \Gamma}{(1+i\omega\tau)\hbar^{2}}\left(\hat{z}\times\tilde{\boldsymbol{\nabla}T}\right)
\left(\hat{z}\times\tilde{\boldsymbol{\nabla}{T}}\right)\cdot \boldsymbol{E}.\\
\end{aligned}
\end{equation}

\setcounter{equation}{0}
\setcounter{figure}{0}
\setcounter{table}{0}
\makeatletter
\renewcommand{\theequation}{C\arabic{equation}}
\renewcommand{\thefigure}{C\arabic{figure}}
\renewcommand{\thetable}{C\arabic{table}}

\bigskip
\bigskip

\noindent
\noindent
\section{Angular dependence of the electric-field-induced nonlinear anomalous Nernst coefficient}{\label{D-app}}
We consider the 2D case with $\mathcal{T}$ symmetry and assume a principal crystal axis along $x$-direction. Then, the nonlinear second harmonic response along $x$ and $y$-direction in this Cartesian coordinate can be written as
\begin{widetext}
\begin{equation}
\begin{aligned}
&\left(
\begin{array}{c}
   j_{x}^{\mathrm{nl},2\omega}\\
   j_{y}^{\mathrm{nl},2\omega}
   \end{array}
   \right)
=
\left(
\begin{array}{cccccccc}
  0             &          0               &           0       &        0   &
 -\xi_{yxxx}       &       -\xi_{yxxy}    &            \xi_{xyyx}        &   \xi_{xyyy}\\
  \xi_{yxxx}    &        \xi_{yxxy}        &        -\xi_{xyyx}       &      -\xi_{xyyy}          &
  0  &        0      &              0               &         0   \\
   \end{array}
   \right)
\left(
\begin{array}{c}
   \tilde{\partial_{x}T}\tilde{\partial_{x}T}E_{x}\\
    \tilde{\partial_{x}T}\tilde{\partial_{x}T}E_{y}\\
    \tilde{\partial_{x}T}\tilde{\partial_{y}T}E_{x}\\
    \tilde{\partial_{x}T}\tilde{\partial_{y}T}E_{y}\\
    \tilde{\partial_{y}T}\tilde{\partial_{x}T}E_{x}\\
    \tilde{\partial_{y}T}\tilde{\partial_{x}T}E_{y}\\
    \tilde{\partial_{y}T}\tilde{\partial_{y}T}E_{x}\\
    \tilde{\partial_{y}T}\tilde{\partial_{y}T}E_{y}\\
   \end{array}
   \right). \\
\end{aligned}
\label{App-D-total}
\end{equation}
\end{widetext}
To obtain Eq.~(\ref{App-D-total}), the relations $\xi_{abcd}=-\xi_{bacd}$ with $a ~\text{and} ~b=\{x,y\}$ [see details in Sec.~\ref{SA-AD}] have been used. When one transforms the principal axis coordinate into the coordinates defined by the measurement setup [Fig.~\ref{figure-sch}], in which $\parallel$ is the direction of the temperature gradient (meaning $\tilde{\partial_{\perp} T}=0$) and $\perp$ represents the Nernst current direction perpendicularly to the temperature gradient, the nonlinear second harmonic current, temperature gradient and electric field transforms as, respectively
\begin{equation}
\begin{aligned}
\left(
\begin{array}{c}
   j_{x}^{\mathrm{nl},2\omega}\\
   j_{y}^{\mathrm{nl},2\omega}
   \end{array}
   \right)
    &=
\left(
\begin{array}{cc}
   \cos\theta            &          -\sin\theta \\
   \sin\theta   &     \cos\theta
   \end{array}
   \right)
 \left(
\begin{array}{c}
   j_{\parallel}^{\mathrm{nl},2\omega} \\
   j_{\perp}^{\mathrm{nl},2\omega}
   \end{array}
   \right),
\end{aligned}
\label{App-D-JF}
\end{equation}
\begin{equation}
\begin{aligned}
\left(
\begin{array}{c}
   \tilde{\partial_{x}T}\\
   \tilde{\partial_{y}T}
   \end{array}
   \right)
    &=
 \left(
\begin{array}{c}
   \cos\theta  \\
   \sin\theta
   \end{array}
   \right) \tilde{\partial T},
\end{aligned}
\label{App-D-delt}
\end{equation}
\begin{equation}
\begin{aligned}
\left(
\begin{array}{c}
  E_{x}\\
  E_{y}
   \end{array}
   \right)
   &=
\left(
\begin{array}{cc}
   \cos\theta            &          -\sin\theta \\
   \sin\theta   &     \cos\theta
   \end{array}
   \right)
 \left(
\begin{array}{c}
    E_{\|}\\
    E_{\bot}
   \end{array}
   \right)
   \\
  & =E
\left(
\begin{array}{cc}
   \cos\theta            &          -\sin\theta \\
   \sin\theta   &     \cos\theta
   \end{array}
   \right)
 \left(
\begin{array}{c}
   \cos(\varphi-\theta) \\
   \sin(\varphi-\theta)
   \end{array}
   \right),
\end{aligned}
\label{App-D-trc}
\end{equation}
where the $\theta$ ($\varphi$) indicates the polar angle of the temperature gradient $\tilde{\boldsymbol{\nabla}T}$ (the dc electric field $\boldsymbol{E}^\text{dc}$) measured from the $x$ direction, and $\tilde{\partial T}=\tilde{\partial_{\parallel} T}=[(\tilde{\partial_{x}{T}})^{2}+(\tilde{\partial_{y}{T}})^{2}]^{1/2}$ indicates the amplitude of the temperature gradient for simplification. 
According to the transform relations of  temperature gradient [Eq. \eqref{App-D-delt}]  and  electric field [Eq.~\eqref{App-D-trc}], we can have
\begin{equation}
\begin{aligned}
\left(
\begin{array}{c}
  \tilde{\partial_{x}T}\tilde{\partial_{x}T}E_{x}\\
    \tilde{\partial_{x}T}\tilde{\partial_{x}T}E_{y}\\
    \tilde{\partial_{x}T}\tilde{\partial_{y}T}E_{x}\\
    \tilde{\partial_{x}T}\tilde{\partial_{y}T}E_{y}\\
    \tilde{\partial_{y}T}\tilde{\partial_{x}T}E_{x}\\
    \tilde{\partial_{y}T}\tilde{\partial_{x}T}E_{y}\\
    \tilde{\partial_{y}T}\tilde{\partial_{y}T}E_{x}\\
    \tilde{\partial_{y}T}\tilde{\partial_{y}T}E_{y}\\
   \end{array}
   \right)
    = \left(
\begin{array}{c}
    \cos^{2}\theta\cos\varphi         \\
   \cos^{2}\theta\sin\varphi        \\
   \sin\theta\cos\theta\cos\varphi   \\
   \sin\theta\cos\theta\sin\varphi    \\
   \sin\theta\cos\theta\cos\varphi   \\
   \sin\theta\cos\theta\sin\varphi   \\
   \sin^{2}\theta\cos\varphi       \\
   \sin^{2}\theta\sin\varphi         \\
   \end{array}
   \right)\tilde{\partial T}\tilde{\partial T} E .
\end{aligned}
\label{App-D-fdff}
\end{equation}
Combining the relations with  Eqs.~(\ref{App-D-fdff})~(\ref{App-D-total}) and (\ref{App-D-JF}), one can obtain
\begin{equation}
\begin{aligned}
j_{\parallel}^{\mathrm{nl},2\omega}&=0,\\
j_{\perp}^{\mathrm{nl},2\omega}&=\xi_\text{N}(\theta,\varphi)\tilde{\partial T}\tilde{\partial T} E,\\
\end{aligned}
\label{Ap-C-6}
\end{equation}
with the angle-dependent scalar coefficient determined by
\begin{equation}
\begin{aligned}
\xi_\text{N}(\theta,\varphi)=
&(\xi_{yxxx}\cos\varphi+\xi_{yxxy}\sin\varphi)\cos\theta \\
&-(\xi_{xyyx}\cos\varphi+\xi_{xyyy}\sin\varphi)\sin\theta. \\
\end{aligned}
\end{equation}

\setcounter{equation}{0}
\setcounter{figure}{0}
\setcounter{table}{0}
\makeatletter
\renewcommand{\theequation}{D\arabic{equation}}
\renewcommand{\thefigure}{D\arabic{figure}}
\renewcommand{\thetable}{D\arabic{table}}

\bigskip
\bigskip

\noindent
\noindent
\section{The angular dependence of the nonlinear Nernst voltage}{\label{app-voltage}}
We have analyzed that angular dependence of the electric-field-induced NANC $\xi_\text{N}(\theta,\varphi)$ in appendix \ref{D-app}. However, what is measured in experiments
is the nonlinear {\color{blue}{Nernst}} voltage, which will be derived in this appendix based on Ohm's law. The detected Nernst second-harmonic voltage $V^{2\omega}_{\perp}$ as a second-order response to alternating temperature gradient via NANE can be written as
\begin{equation}
V^{2\omega}_{\perp}=WE^{2\omega}_{\perp},
\label{App-D-votlage}
\end{equation}
where $W$ indicates the width of sample vertical to the temperature gradient, and $E^{2\omega}_{\perp}$ represents the induced electric field perpendicular to temperature gradient owing to the NANE, which can be determined through the generated second-harmonic currents $j_{\parallel}^{\mathrm{nl},2\omega}$ ($j_{\perp}^{\mathrm{nl},2\omega}$) parallel (perpendicular) to temperature gradient as follows. When transforming the coordinates defined by the measurement setup [Fig.~\ref{figure-sch}] into the principal coordinate (having assumed a principal crystal direction along $x$ axis) , we have
\begin{equation}
\begin{aligned}
\left(
\begin{array}{c}
   j_{\parallel}^{\mathrm{nl},2\omega}\\
   j_{\perp}^{\mathrm{nl},2\omega}
   \end{array}
   \right)
    &=
\left(
\begin{array}{cc}
   \cos\theta            &\sin\theta \\
   -\sin\theta   &     \cos\theta
   \end{array}
   \right)
 \left(
\begin{array}{c}
  j_{x}^{\mathrm{nl},2\omega} \\
  j_{y}^{\mathrm{nl},2\omega}
   \end{array}
   \right),
\end{aligned}
\label{App-fd-JF}
\end{equation}
and
\begin{equation}
\begin{aligned}
\left(
\begin{array}{c}
   E^{2\omega}_{\parallel}\\
   E^{2\omega}_{\perp}
   \end{array}
   \right)
    &=
\left(
\begin{array}{cc}
   \cos\theta    &\sin\theta \\
   -\sin\theta   &\cos\theta
   \end{array}
   \right)
 \left(
\begin{array}{c}
   E^{2\omega}_{x} \\
   E^{2\omega}_{y}
   \end{array}
   \right),
\end{aligned}
\label{App-D-electric}
\end{equation}
In addition, according to Ohm's law, we have the relations between conductivity $\sigma$ and the second-harmonic current $j_{x}^{\mathrm{nl},2\omega}$ ($j_{y}^{\mathrm{nl},2\omega}$), which is  parallel (perpendicular) to the principal axis (i.e. $x$ axis) of the system,  as $j_{x}^{\mathrm{nl},2\omega}=\sigma_{xx} E^{2\omega}_{x}$ ($j_{y}^{\mathrm{nl},2\omega}=\sigma_{yy} E^{2\omega}_{y}$). The combination of these relation with Eqs.~\eqref{App-fd-JF} and \eqref{App-D-electric} yields
\begin{equation}
\begin{aligned}
 &E^{2\omega}_{\perp}=j_{\perp}^{\mathrm{nl},2\omega}\left(\frac{\sin^{2}\theta}{\sigma_{xx}}+
 \frac{\cos^{2}\theta}{\sigma_{yy}}\right),\\
 &E^{2\omega}_{\parallel}=j_{\perp}^{\mathrm{nl},2\omega}\left(\frac{1}{\sigma_{yy}}-\frac{1}
 {\sigma_{xx}}\right)\sin\theta\cos\theta.\\
 \end{aligned}
\label{relations}
\end{equation}
Combining Eq.(\ref{relations}) with Eq.(\ref{App-D-votlage}) and assuming $\sigma_{xx}=\sigma_{yy}$, the second-harmonic Nernst voltage $V^{2\omega}_{\perp}$ can be expressed as
\begin{equation}
V^{2\omega}_{\perp}=\frac{Wj_{\perp}^{\mathrm{nl},2\omega}}{\sigma_{xx}}.
\end{equation}
The $j_{\perp}^{\mathrm{nl},2\omega}$, perpendicular to the temperature gradient, indicate the second-harmonic component of the nonlinear Nernst current and, therefore, has been written as  $j_{\mathrm{N}}^{\mathrm{nl},2\omega}$ in the main text.

\setcounter{equation}{0}
\setcounter{figure}{0}
\setcounter{table}{0}
\makeatletter
\renewcommand{\theequation}{E\arabic{equation}}
\renewcommand{\thefigure}{E\arabic{figure}}
\renewcommand{\thetable}{E\arabic{table}}

\bigskip
\bigskip

\noindent
\noindent
\section{The experimental isolation from other potential signals}
\label{EIFO}
When applying an alternating temperature gradient $\boldsymbol{\nabla} {T}^{\omega}(t)=\text{Re}\{ \tilde{\boldsymbol{\nabla} T} e^{i\omega t}\}$ and an external dc electric field $\boldsymbol{E}^\text{dc}$ to the sample, the output voltage $V_\text{out}$ generated from the $\boldsymbol{\nabla} {T}^{\omega}(t)$ can be expressed as
\begin{equation}
V_\text{out}=\text{Re}[V^{0}+V^{\omega}e^{i\omega t}+V^{2\omega}e^{i2\omega t}+V^{3\omega}e^{3i\omega t}+...],
\label{App-E-df}
\end{equation}
where $\text{Re}\{V^{0}\}$ and $\text{Re}\{V^{n\omega}\}$ represent the amplitude of the direct and nth-harmonic voltage, respectively.

By extracting the second-harmonic voltage $V^{2\omega}$ via lock-in amplification at $2\omega$ frequency, one can selectively detect the nonlinear Nernst signal as a second-order response to temperature gradient.
This phase-sensitive detection at $2\omega$ inherently rejects all signals that do not oscillate at $2\omega$. Consequently, the second-harmonic detection inherently excludes (a) linear/ nonlinear Hall signals, which are frequency independent; (b)non-second-order anomalous Nernst effects, which either appear as dc signal or oscillate at frequencies other than $2\omega$. {\color{blue}{Therefore, the lock-in amplification detection at $2\omega$ inherently isolates the response quadratic in the temperature gradient. Next, we will discuss how to distinguish the BCP contribution from other second-harmonic signals through field, angle, and scaling approaches.
}}

In addition to the proposed electric-field-induced NANE originating from BCP, other quantum-geometric contributions to the second-harmonic voltage signals may also exist: (i) relaxation-time-independent intrinsic NANE originating from the orbital magnetic quadrupole moment or thermal Berry connection polarizability; (ii)${\boldsymbol{\Lambda}}^{T}$-induced inherent NANE, which is independent of $\boldsymbol{E}^\text{dc}$; (iii) the electric-induced NANE from the anomalous velocity contribution ($\xi^\text{NDF-AV}_{abcd}$). In $\mathcal{T}$ symmetric systems considered in our works, the intrinsic NANE is forbidden. The other two types vanish when both
 $\mathcal{T}$ and $\mathcal{P}$ symmetries are present.

 For the $\mathcal{P}$ symmetry broken systems, 
 one can further exclude the ${\boldsymbol{\Lambda}}^{T}$-induced inherent NANE from the electric-field-induced NANE via field- and  angle-dependence measurements: measuring $V^{2\omega}$ versus $E^\text{dc}$ with all other parameters fixed yields background at $E^\text{dc}=0$, which corresponds to the parasitic ${\boldsymbol{\Lambda}}^{T}$-induced signal. Rotating $\boldsymbol{E}^\text{dc}$ while keeping temperature gradient fixed leaves the background signal unchanged (angle independence), whereas the proposed BCP signal depends on $\boldsymbol{E}^\text{dc}$ orientation, for instance, a sinusoidal dependence in monolayer graphene [Eq.~\eqref{D-1}]. Hence, subtracting the angle-independent background isolates the BCP contribution. As discussed in Sec. \ref{SA-AD}, the anomalous-velocity contribution in $\mathcal{P}$-symmetric systems can be suppressed by aligning  $\boldsymbol{E}^\text{dc}$ parallel or antiparallel to the temperature gradient.

The extrinsic contributions (side jump and skew scattering) from the disorder may also contribute to the second-harmonic Nernst voltage.  However, side-jump contribution has been proved either to vanish in NANE under $C_{3}$ symmetry \cite{Liu-2025} or to be negligible in high-mobility systems \cite{Liu-2025,Zhao-2023,Nagaosa-2010}, leaving skew-scattering contribution as the residual spurious signal.  The BCP contribution ($\sim\tau$) can be distinguished from the skew-scattering contribution ($\sim\tau^{3}$) using the established scaling laws between the NANE and linear Seebeck coefficient $S_{\parallel}$ ($\sim\tau$), \cite{Liu-2025} namely
\begin{equation}
\frac{V_{\perp}^{2\omega}}{V_{\parallel}^{2}}=A(\sigma_{\parallel}S_{\parallel})^{2}+B,
\label{App-E-V}
\end{equation}
where $V_{\perp}^{2\omega}$ ($V_{\parallel}$) denotes the generated second-harmonic (first-harmonic) voltage perpendicular (aligned) to temperature gradient, $\sigma_{\parallel}$ is the longitudinal conductivity. The scaling parameters $A$ arises from skew-scattering contribution while $B$ comes from BCP contribution. Notably, if $C_{3}$ symmetry is broken, $B$ includes contributions from both BCP and side-jump mechanisms.


\begin{thebibliography}{99}

\bibitem{Z. Qiao2010} Z. Qiao, S. A. Yang, W. Feng, W.-K. Tse, J. Ding, Y. Yao, J. Wang, and Q. Niu, Quantum anomalous Hall effect in graphene from Rashba and exchange effects, Phys. Rev. B \textbf{82}, 161414(R) (2010).

\bibitem{G. Xu2011} G. Xu, H. Weng, Z. Wang, X. Dai, and Z. Fang, Chern semimetal and the quantized anomalous Hall effect in $\mathrm{HgCr_{2}Se_{4}}$, Phys. Rev. Lett. \textbf{107}, 186806 (2011).

\bibitem{D. Xiao2006} D. Xiao, Y. Yao, Z. Fang, and Q. Niu, Berry-phase effect in anomalous thermoelectric transport, Phys. Rev. Lett. \textbf{97}, 026603 (2006).

\bibitem{Uchida-2021} K. I. Uchida, W. Zhou, Y. Sakuraba,Transverse thermoelectric
generation using magnetic materials. Appl. Phys. Lett. \textbf{118}, 140504(2021).

\bibitem{Uchida-2022} K. I. Uchida, J. P. Heremans, Thermoelectrics: from longitudinal
to transverse. Joule \textbf{6}, 2240¨C2245 (2022).


\bibitem{I. Sodemann2015} I. Sodemann and L. Fu, Quantum nonlinear Hall effect induced by Berry curvature dipole in time-reversal invariant materials, Phys. Rev. Lett. \textbf{115}, 216806 (2015).

\bibitem{T.Low2015} T. Low, Y. Jiang, and F. Guinea, Topological currents in black phosphorus with broken inversion symmetry, Phys. Rev. B \textbf{92}, 235447 (2015).

\bibitem{Z.Z.Du2018} Z. Z. Du, C. M. Wang, H.-Z. Lu, and X. C. Xie, Band signatures for strong nonlinear Hall effect in bilayer $\mathrm{WTe_{2}}$, Phys. Rev. Lett. \textbf{121}, 266601 (2018).

\bibitem{Z.Du2019} Z. Du, C. Wang, S. Li, H.-Z. Lu, and X. Xie, Disorderinduced nonlinear Hall effect with time-reversal symmetry, Nat. Commun. \textbf{10}, 3047 (2019).

\bibitem{R.Battilomo2019} R. Battilomo, N. Scopigno, and C. Ortix, Berry curvature dipole in strained graphene: A Fermi surface warping effect, Phys. Rev. Lett. \textbf{123}, 196403 (2019).

\bibitem{J.I.Facio2018} J. I. Facio, D. Efremov, K. Koepernik, J.-S. You, I. Sodemann, and J. van den Brink, Strongly enhanced Berry dipole at topological phase transitions in $\mathrm{BiTeI}$, Phys. Rev. Lett. \textbf{121}, 246403 (2018).

\bibitem{Z.Du2021} Z. Du, C. Wang, H.-P. Sun, H.-Z. Lu, and X. Xie, Quantum theory of the nonlinear Hall effect, Nat. Commun. \textbf{12}, 5038 (2021).

\bibitem{A.Bandyopadhyay2024} A. Bandyopadhyay, N. B. Joseph, and A. Narayan, Non-linear Hall effects: Mechanisms and materials, Mater. Today Electron. \textbf{8} 100101 (2024).

\bibitem{W.Miao2023} W. Miao, W.-L. Zhen, C. Tan, J. Wang, Y. Nie, H. Wang, L. Wang, Q. Niu and M.-L. Tian, Nonreciprocal antisymmetric magnetoresistance and unconventional Hall effect in a two-dimensional ferromagnet, ACS Nano \textbf{17}, 25449 (2023).

\bibitem{D. Kaplan2024} D. Kaplan, T. Holder, and B. Yan, Unification of nonlinear anomalous Hall effect and nonreciprocal magnetoresistance in metals by the quantum geometry, Phys. Rev. Lett. \textbf{132}, 026301 (2024).

\bibitem{V.A.Zyuzin2020} V. A. Zyuzin, In-plane hall effect in two-dimensional helical electron systems, Phys. Rev. B \textbf{102}, 241105(R) (2020).

\bibitem{Hui Wang2024} H. Wang, Y.-X. Huang, H. Liu, X. Feng, J. Zhu, W. Wu, C. Xiao, and S. A. Yang, Orbital origin of the intrinsic planar Hall effect, Phys. Rev. Lett. \textbf{132}, 056301 (2024).

\bibitem{C. Zeng2020} C. Zeng, S. Nandy, and S. Tewari, Fundamental relations for anomalous thermoelectric transport coefficients in the nonlinear regime, Phys. Rev. Res. \textbf{2}, 032066(R) (2020).

\bibitem{D.-K. Zhou2022} D.-K. Zhou, Z.-F. Zhang, X.-Q. Yu, Z.-G. Zhu, and G. Su, Fundamental distinction between intrinsic and extrinsic nonlinear thermal Hall effects, Phys. Rev. B \textbf{105}, L201103 (2022).

\bibitem{H.Varshney2} H. Varshney, K. Das, P. Bhalla, and A. Agarwal, Quantum kinetic theory of nonlinear thermal current, Phys. Rev. B \textbf{107}, 235419 (2023).

\bibitem{H. Varshney2023} H. Varshney, R. Mukherjee, A. Kundu, and A. Agarwal, Intrinsic nonlinear thermal Hall transport of magnons: A quantum kinetic theory approach, Phys. Rev. B \textbf{108}, 165412 (2023).

\bibitem{Zhang-2025} Y.-F. Zhang, Z.-F. Zhang, Z.-G. Zhu, and G. Su, Second-order intrinsic Wiedemann-Franz law, Phys. Rev. B \textbf{111}, 165424(2025).


\bibitem{X.-Q. Yu2019}X.-Q. Yu, Z.-G. Zhu, J.-S. You, T. Low, and G. Su, Topological nonlinear anomalous Nernst effect in strained transition metal dichalcogenides, Phys. Rev. B \textbf{99}, 201410(R) (2019).

\bibitem{C. Zeng2019} C. Zeng, S. Nandy, A. Taraphder, and S. Tewari, Nonlinear Nernst effect in bilayer $\mathrm{WTe_{2}}$, Phys. Rev. B \textbf{100}, 245102 (2019).

\bibitem{Robert-2005}  Robert E. Newnham, Properties of Materials: Anisotropy,
Symmetry, Structure (Oxford University Press, New York,2005).

\bibitem{Y.-L. Wu2021} Y.-L. Wu, G.-H. Zhu, and X.-Q. Yu, Nonlinear anomalous nernst effect in strained graphene induced by trigonal warping, Phys. Rev. B \textbf{104}, 195427 (2021).

\bibitem{Harsh Varshney2025} H. Varshney and A. Agarwal, Intrinsic nonlinear Nernst and Seebeck effect, New J. Phys. \textbf{27}, 083506 (2025).

\bibitem{Liu-2025} H. Liu, H. Jiang, J. Li, Z. Zhang, X. C. Xie, P. He, J. Zhai, M. Zhang, and J. Shen, Nonlinear Nernst effect in trilayer graphene at zero magnetic field, Nat. Nanotech. \textbf{20}, 1221-1227 (2025).

\bibitem{Y. Gao2018} Y. Gao and D. Xiao, Orbital magnetic quadrupole moment and nonlinear anomalous thermoelectric transport, Phys. Rev. B \textbf{98}, 060402(R) (2018).


\bibitem{X.-G. Ye2023} X.-G. Ye, H. Liu, P.-F. Zhu, W.-Z. Xu, S. A. Yang, N. Shang, K. Liu, and Z.-M. Liao, Control over Berry curvature dipole with electric field in $\mathrm{WTe}_{2}$, Phys. Rev. Lett. \textbf{130}, 016301 (2023).

\bibitem{A. Bhattacharya2025} A. Bhattacharya and A. M. Black-Schaffer, Electric field induced second-order anomalous hall transport in unconventional Rashba systems, Phys. Rev. B \textbf{111}, L041202 (2025).

\bibitem{A.Mukherjee2025} A. Mukherjee, B. Sanyal, A. M. Black-Schaffer, and A. Bhattacharya, Electric field controlled second-order anomalous hall effect in altermagnets, arXiv:2510.14899.

\bibitem{S. Korrapati2025} S. Korrapati, S. Nandy and S. Tewari, Electric field induced Berry curvature dipole and non-linear anomalous Hall effect in higher wave symmetric unconventional magnets, arXiv:2510.20237v1.

\bibitem{H.Li2023} H. Li, M. Li, R.-C. Xiao, W. Liu, L. Wu, W. Gan, H. Han, X. Tang, C. Zhang, and J. Wang, Current induced second-order nonlinear Hall effect in bulk $\mathrm{WTe_{2}}$, Appl. Phys. Lett. \textbf{123}(16), 163102 (2023).

\bibitem{J.Yang2025}  J. Yang, L. Wei, Y. Li, L. Chen, W. Niu, S. Wang, F. Li, P. Liu, S. Zhou, and Y. Pu, Electric field control of nonlinear Hall effect in the type-II Weyl semimetal $\mathrm{TaIrTe_{4}}$, Appl. Phys. Lett. \textbf{127}, 033102 (2025).

\bibitem{Tanay-2021} T. Nag, S. K. Das, C. Zeng, and S. Nandy, Third-order Hall effect in the surface states of a topological insulator, Phys. Rev. B \textbf{107}, 245141(2023).

\bibitem{Y. Gao2014} Y. Gao, S. A. Yang, and Q. Niu, Field-induced positional shift of Bloch electrons and its dynamical implications, Phys. Rev. Lett. \textbf{112}, 166601 (2014).

\bibitem{S. Lai2021} S. Lai, H. Liu, Z. Zhang, J. Zhao, X. Feng, N. Wang, C. Tang, Y. Liu, K. S. Novoselov, S. A. Yang, and W.-B. Gao, Third-order nonlinear Hall effect induced by the Berry-connection polarizability tensor, Nat. Nanotechnol. \textbf{16}, 869 (2021).

\bibitem{H.Liu2021} H. Liu, J. Zhao, Y.-X. Huang, W. Wu, X.-L. Sheng, C. Xiao, and S. A. Yang, Intrinsic second-order anomalous hall effect and its application in compensated antiferromagnets, Phys. Rev. Lett.
\textbf{127}, 277202 (2021).

\bibitem{H. Liu2022} H. Liu. J. Zhao, Y.-X. Huang, X. Feng, C. Xiao, W. Wu, S. Lai, W.-B. Gao, and S. A. Yang, Berry connection polarizability tensor and third-order Hall effect, Phys. Rev. B \textbf{105}, 045118 (2022).



\bibitem{D.Xiao2007}  D. Xiao, W. Yao, and Q. Niu, Valley-contrasting physics in graphene: Magnetic moment and topological transport, Phys. Rev. Lett. \textbf{99}, 236809 (2007).

\bibitem{Castro-Neto2009} A. H. Castro Neto, F. Guinea, N. M. R. Peres, K. S. Novoselov, and A. K. Geim, The electronic properties of graphene, Rev. Mod. Phys. \textbf{81}, 109 (2009).

\bibitem{Chester-1963} G. V. Chester, The Boltzmann equation for inhomogeneous electric fields, Proc. Phys. Soc. \textbf{81}, 938 (1963).

\bibitem{Pottier-2010} N. Pottier, Nonequilibrium Statistical Physics: Linear Irreversible Processes, Oxford University Press, Oxford, 2010.

\bibitem{Adam-2007}S. Adam, E. H. Hwang, V. M. Galitski, and S. Das Sarma, A self-consistent theory for graphene transport,
  {\color{cyan}{PNAS}} \textbf{104}, 18392 (2007).

\bibitem{Deal-2010} C. R. Dean, A. F. Young, I. Meric, C. Lee, L. Wang, S. Sorgenfrei, K. Watanabe, T. Taniguchi, P. Kim, K. L. Shepard  and J. Hone, Boron nitride substrates for high-quality graphene electronics, Nat. Nanotechnol. \textbf{5}, 722-726 (2010).

\bibitem{Sekine-2020} A. Sekine and N. Nagaosa, Quantum kinetic theory of thermoelectric and thermal transport in a magnetic field, Phys. Rev. B \textbf{101}, 155204 (2020).

\bibitem{Pankaj-2020} P. Bhalla, Intrinsic contribution to nonlinear thermoelectric effects in topological insulators, Phys. Rev. B \textbf{103}, 115304 (2021).


\bibitem{Mayorov-2012} A. S. Mayorov, D. C. Elias, I. S. Mukhin, S. V. Morovov, L. A. Ponomarenko, K. S. Novoselov, A. K. Geim, and R. V. Gorbachev, How close can one approach the Dirac point in Graphene experimentally, Nano. Lett. \textbf{12}, 4629-4634 (2012).

\bibitem{Sarma-2013}S. Das Sarma and  E. H. Hwang, Density dependent electrical conductivity in suspended graphene approaching the Dirac point in transport, Phys. Rev. B \textbf{87}, 035415 (2013).


\bibitem{WU-2025} Ying-Li Wu, Jia-Liang Wan, and Xiao-Qin Yu, Intrinsic nonlinear valley Nernst effect in the strained bilayer graphene, Phys. Rev. B \textbf{112}, 144104 (2025).

\bibitem{Gao-2019} Y. Gao, Semiclassical dynamics and nonlinear charge current, Frontiers of Physics, \textbf{14}(3), 33404 (2019).

\bibitem{Zhao-2023} T.-Y. Zhao, A.-Q. Wang, X.-G. Ye, X.-Y. Liu, X. Liao, and Z.-M. Liao, Gate-tunable Berry curvature dipole polarizability in Dirac semimetal $\mathrm{Cd_{3}As_{2}}$, Phys. Rev. Lett. \textbf{131}, 186302 (2023).

\bibitem{Hwang-2007}E. H. Hwang, S. Adam, and S. Das Sarma, Carrier transport in two-dimensional graphene layers, Phys. Rev. Lett. \textbf{98}, 186806 (2007).

\bibitem{Nagaosa-2010} N. Nagaosa, J. Sinova, S. Onoda, A.H. MacDonald, and N. P. Ong, Anomalous Hall effect, Rev. Mod. Phys. \textbf{82}, 1539 (2010).

{\color{blue}{\bibitem{Uchida} K. Uchida, S. Takahashi, K. Harii, J. Ieda, W. Koshihae, K. Ando, S. Maekawa, and E. Saitoh, Observation of the spin Seebeck effect, Nature (London) \textbf{455}, 778 (2008).}}

{\color{blue}{\bibitem{Aniru-2019} A. Pan and D. C. Marinescu, Nonlinear spin-current generation in quantum wells with arbitrary Rashba-Dresselhaus spin-orbit interactions, Phys. Rev. B \textbf{99}, 245204(2019).}}

{\color{blue}{\bibitem{Kapri-2025} P. Kapri, spin currents in Rashba altermagnets: from equilibrium to nonlinear regimes, Phy. Rev. B \textbf{112}, 155422 (2025).}}

\end{thebibliography}
\end{document}